\documentclass[superscriptaddress, nofootinbib, prd, 10pt]{revtex4}

\usepackage{latexsym, amssymb, amsmath, graphics, graphicx}
\usepackage{verbatim}
\usepackage[colorlinks=true,  linkcolor=black, citecolor=blue, urlcolor=blue]{hyperref}
\usepackage[usenames,dvipsnames]{color}
\usepackage{multirow}
                                    
\newcommand{\GeV}{\text{GeV}}
\newcommand{\MeV}{\text{MeV}}
\newcommand{\eV}{\text{eV}}
\newcommand{\cm}{\text{cm}}

\newcommand{\vev}{\langle h \rangle }

\newcommand{\ministack}[2]{\begin{tabular}{c} {#1} \\[-2pt] {#2} \end{tabular}}

\newcommand{\myparagraph}[1]{\vspace{10pt} {\bf #1} \,\,}

\newcommand{\myref}[2]{\hyperref[#2]{#1~\ref*{#2}}}

\def\simgt{\mathrel{\lower2.5pt\vbox{\lineskip=0pt\baselineskip=0pt
           \hbox{$>$}\hbox{$\sim$}}}}
\def\simlt{\mathrel{\lower2.5pt\vbox{\lineskip=0pt\baselineskip=0pt
           \hbox{$<$}\hbox{$\sim$}}}}

\graphicspath{{figs/}}

\begin{document}

\title{Dark Matter Direct Detection with Accelerometers}

\author{Peter W.~Graham}
\affiliation{Stanford Institute for Theoretical Physics, Department of Physics, Stanford University, Stanford, CA 94305}

\author{David E.~Kaplan}
\affiliation{Stanford Institute for Theoretical Physics, Department of Physics, Stanford University, Stanford, CA 94305}
\affiliation{Department of Physics \& Astronomy, The Johns Hopkins University, Baltimore, MD  21218}
\affiliation{Berkeley Center for Theoretical Physics, Department of Physics, University of California, Berkeley, CA 94720}
\affiliation{Kavli Institute for the Physics and Mathematics of the Universe (WPI), Todai Institutes for Advanced Study, University of Tokyo, Kashiwa 277-8583, Japan}

\author{Jeremy Mardon}
\affiliation{Stanford Institute for Theoretical Physics, Department of Physics, Stanford University, Stanford, CA 94305}

\author{Surjeet Rajendran}
\affiliation{Berkeley Center for Theoretical Physics, Department of Physics, University of California, Berkeley, CA 94720}

\author{William A. Terrano}
\affiliation{Center for Experimental Nuclear Physics and Astrophysics, University of Washington, Seattle, WA 98195-4290}
\affiliation{Physikdepartment, Technische Universit\"at M\"unchen, D-85748 Garching, Germany}

\begin{abstract}
The mass of the dark matter particle is unknown, and may be as low as $\sim$$10^{-22}$\,eV.  The lighter part of this range, below $\sim$\,eV, is relatively unexplored both theoretically and experimentally but contains an array of natural dark matter candidates.
An example is the relaxion, a light boson predicted by cosmological solutions to the hierarchy problem. 
One of the few generic signals such light dark matter can produce is a time-oscillating, EP-violating force.
We propose searches for this using accelerometers, and consider in detail the examples of torsion balances, atom interferometry, and pulsar timing.  
These approaches have the potential to probe large parts of unexplored parameter space in the next several years.  
Thus such accelerometers provide radically new avenues for the direct detection of dark matter.
\end{abstract}

\maketitle

\tableofcontents

\hypersetup{linkcolor=blue} 

\section{Introduction}

The astrophysical and cosmological evidence for dark matter is overwhelming.  
However, we know almost nothing about its fundamental properties. 
All we can say confidently about its mass is that it lies between $\sim\! 10^{-22}\,\eV$~\cite{Khlopov:1985jw, Press:1989id, Sahni:1999qe, Hu:2000ke} and astrophysically large scales.  
The type of particle and the nature of its interactions are unknown. 
Experimentally testing the vast range of well motivated dark matter candidates is one of the most important  objectives of modern physics. 

For the purposes of direct detection, dark matter candidates can be roughly divided into two classes, particle-like or field-like.  Because the local energy density of dark matter is $\rho_\text{DM} \!\sim\! (0.04 \, \eV)^4$, if the dark matter mass (really momentum) is much greater than $\sim\!0.1$\,eV the phase space density will be low and the dark matter acts more like a particle for the purposes of detection.  If the mass (momentum) is much below this scale, then dark matter has a high phase space density (many particles per de Broglie wavelength cubed) and is often well-described as a classical field\footnote{In this case of course it must be a boson.}.  
The Weakly Interacting Massive Particle (WIMP)~\cite{Goodman:1984dc} is the prototypical example of particle-like dark matter, while the axion~\cite{Wilczek:1977pj, Weinberg:1977ma, Peccei:1977hh, Dine:1982ah, Preskill:1982cy, Abbott:1982af} is the prototypical example of field-like dark matter.  
Traditional particle detection techniques are likely the optimal way to search for particle-like dark matter, at least in a very broad range of masses around the Weak scale.  
There has been a significant, decades-long effort in direct detection of dark matter, focused on the WIMP.  While the WIMP is well-motivated, the lack of evidence for it to date at either direct detection experiments or the LHC suggests that the search for dark matter should be broadened to include other candidates.  
This paper will focus on novel ways to search for light, field-like dark matter.  Such light dark matter has attracted a great deal of interest recently both theoretically and experimentally~\cite{Sikivie:1983ip, Asztalos:2009yp, Asztalos:2011bm, Arvanitaki:2009fg, Graham:2011qk, Graham:2013gfa, Budker:2013hfa, Arvanitaki:2014dfa, Kim:2015yna, Marsh:2015xka, Ahlers:2007rd, Jaeckel:2007ch, Povey:2010hs, Nelson:2011sf, Arias:2012az, Horns:2012jf, An:2013yua, Parker:2013fba, Betz:2013dza, Graham:2014sha,   Arias:2014ela, Chaudhuri:2014dla, An:2014twa, Graham:2015rva, Pospelov:2012mt, Derevianko:2013oaa, Arvanitaki:2014faa, Arvanitaki:2015iga, Izaguirre:2014cza,  Stadnik:2013raa, Stadnik:2014xja, Stadnik:2014tta, Stadnik:2015xbn, Arvanitaki:2010sy, Pani:2012vp, Arvanitaki:2014wva, Brito:2015oca}.

We wish to broaden the search for dark matter to the lighter part of the allowed range, but of course with an additional focus on the  candidates that have the best motivation.  
Part of the motivation for both WIMPs and axions is that they can arise from solutions to tuning problems---the hierarchy problem and the strong CP problem respectively.  
In fact, the recently proposed cosmological solution to the hierarchy problem~\cite{Graham:2015cka} predicts a light field, the relaxion---a light scalar which couples to matter through the Higgs portal.  
This is one of the types of particle that we focus on finding ways to search for in this paper.  
The other strong part of the motivation for WIMPs and axions is that they are good dark matter candidates, meaning that they have natural production mechanisms and are simple effective field theories which describe broad classes of higher-energy models.  
There are several types of light bosons which are also natural dark matter candidates.  
In particular, there are natural production mechanisms for light bosons: misalignment production for scalars~\cite{Dine:1982ah, Preskill:1982cy} and inflationary fluctuation production for vectors~\cite{Graham:2015rva}.  
If a light boson does exist in the theory, these production mechanisms necessarily produce an abundance of this particle in the universe, and it is natural for it to make up some or all of the dark matter.

Several direct detection searches for axion and hidden-photon dark matter are operating or are under construction~\cite{Asztalos:2009yp, Asztalos:2011bm, Budker:2013hfa, Chaudhuri:2014dla, Horns:2014qta, Suzuki:2015sza}.  
In this paper we focus on two other well-motivated possibilities: scalars coupled to the Standard Model through the Higgs portal~\cite{Piazza:2010ye, Arvanitaki:2014faa} and $B$$-$$L$ coupled vectors.  
As we show, such dark matter candidates can be powerfully probed with precision accelerometers---in particular through the time-oscillating, equivalence principle violating force that they exert on normal matter. 
Such accelerometers have been used to search for new forces or modifications of general relativity. 
Here we are considering the possibility that such a new force carrier is itself also the dark matter. 
Interestingly, the direct effect of the field as dark matter is expected to be significantly larger than its effect as a 5th-force mediator in future searches.

\section{Executive Summary}

\begin{itemize}

\item We investigate dark matter in the large mass range below $\sim$\,eV.  Here dark matter behaves as an oscillating classical field.  
In terms of low energy physics, there are only a handful of possible effective theories for dark matter in this part of parameter space, classified in Table \ref{table:bosons} (see \myref{Section}{sec: theory}).  
And in fact there are only four rough classes of detectable effects of such dark matter: spin effects, electromagnetic effects, accelerations, and variations of fundamental constants.

\item Direct detection experiments have already been designed to search for the spin and electromagnetic couplings (e.g.~axion detectors).  Here we focus on the equivalence-principle (EP) violating acceleration caused by dark matter, which is not currently being searched for in any experiment.  This effect is generic for both scalar and vector DM, two natural candidates being a scalar coupled through the Higgs portal and a vector coupled to $B-L$ charge.  In addition we point out a new way of seeing the variation of fundamental constants using pulsar timing arrays.

\item
This DM-induced acceleration oscillates in time with a fixed frequency and long coherence time ($\simgt\!10^6$ periods) and points in a random direction (fixed in the galactic frame over a coherence time). 
These features distinguish it from backgrounds such as seismic noise. 
The amplitude of the acceleration, combined with its distinctive features, make the DM signal significantly easier to search for over a wide DM mass range than static EP-violation (caused by the same light field sourced by the Earth).  See \myref{Section}{sec: ep signature}.

\item 
Existing accelerometer technology designed to search for static EP-violation is automatically also sensitive to a DM-induced signal.
In \myref{Section}{sec:torsion-balance} we examine the potential sensitivity of torsion pendulum setups, for example as used by the E\"ot-Wash group. 
These benefit significantly from the random direction of the DM signal, which avoids the $\sim\!10^{-3}$ suppression suffered in searches for a vertical acceleration sourced by the Earth. 
The impressive sensitivity of these setups means that even a reanalysis of existing data will be able to probe new DM parameter space. 
With the technology upgrades expected over the next several years, they will reach into unconstrained parameter space by many orders of magnitude in mass and coupling strength.

\item
Atom interferometers are another type of high-precision accelerometer which can be sensitive to light field DM. 
As we discuss in \myref{Section}{sec:atom-int}, these are also expected to reach deep into currently unconstrained parameter space over the next several years.

\item
Two other potentially powerful experimental options are lunar laser ranging and pulsar timing arrays, see \myref{Sections}{sec: LLR} \&~\ref{Sec:Pulsars}. 
In both cases, a reanalysis of existing data should be able to constrain new DM parameter space or even uncover a signal, and expected upgrades over the next few years should improve their reach further. 
Pulsar timing arrays in particular can be the most powerful probe of scalar DM at the lowest frequencies.  

\item
Our main results are shown in Figs.~\ref{fig:B-L-plot},~\ref{fig:scalar-plot} \&~\ref{fig:scalar-plot-2}, 
respectively for a $B-L$ coupled vector, a scalar coupled through the Higgs portal, and a scalar coupled to the electron mass operator. 
For the latter scalar coupling, a non fine-tuned part of parameter space appears reachable in the near to mid term, while for the former scalar coupling this looks extremely difficult. 
However in both cases there is a large reach into fine-tuned but otherwise unconstrained parameter space. 
The projected reach for the $B-L$ coupled vector is particularly striking, extending many orders of magnitude into unprobed parameter space.

\end{itemize}

\section{Theoretical Landscape of Light Dark Matter}
\label{sec: theory}

In this section, for theoretically minded readers, we categorize the range of possible light field dark matter candidates and their leading physical effects.

Light bosons are a well-motivated and under-explored class of dark matter candidates.  
While the allowed mass range is vast, the types of fields and natural couplings are limited.  
The leading interactions are shown in Table~\ref{table:bosons}.  
We focus on these operators since operators of higher dimension with linear couplings will have a lesser impact and will generate these lower dimension operators via loops.  
In addition, we ignore operators with quadratic couplings (such as $\phi^2 h^\dagger h$), as their impact will be significantly smaller. 
We postpone discussion of spin-2 candidates for future work, since there are questions about the range of validity, degree of fine tuning, and allowed interactions of massive spin 2 fields (see e.g.~\cite{ArkaniHamed:2002sp, deRham:2014zqa, Dubovsky:2004ud}). 
Finally, we do not consider massive bosons of spin 3 or higher, since we know of no effective field theory valid parametrically above their mass.

\begin{table}[h]
\vspace{10pt}
  \begin{tabular}{|c|c|c|c|c|c|}\hline 
Spin & Type & Operator & Interaction & Oscillating DM Effects & Searches \\\hline 
\multirow{5}{*}{0} & \multirow{2}{*}{scalar} &  \multirow{2}{*}{$ \phi \, h^\dagger h$, $\phi \, \mathcal O_{\rm SM}$} & \multirow{2}{*}{Higgs portal / dilaton} & $m_e, m_p, \alpha$ variation& Atomic clocks~\cite{VanTilburg:2015oza} \\\cline{5-6} 
   &  &  &  & acceleration & $\star$ \\\cline{2-6}  
   & \multirow{3}{*}{pseudo-scalar} & $a \, G^{\mu\nu}\tilde{G}_{\mu\nu}$ & axion-QCD & nucleon EDM & CASPEr~\cite{Budker:2013hfa} \\\cline{3-6}
   &  & ${a}\, F^{\mu\nu}\tilde{F}_{\mu\nu}$ & axion-E\&M & EMF along $B$ field & ADMX~\cite{Asztalos:2009yp} \\\cline{3-6}
   &  & $({\partial_\mu a}) \bar{\psi}\gamma^\mu \gamma_5 \psi$ & axion-fermion &  spin torque & CASPEr~\cite{Budker:2013hfa} \\\hline
\multirow{4}{*}{1} & \multirow{3}{*}{vector} & $A_\mu' \bar{\psi} \gamma^\mu \psi$ & minimally coupled  & acceleration & $\star$ \\\cline{3-6}  
   &  & $F_{\mu\nu}' F^{\mu\nu}$ & vector--photon mixing & EMF in vacuum & DM Radio~\cite{Chaudhuri:2014dla}, ADMX \\\cline{3-6}
   &  & $F_{\mu\nu}' \bar{\psi}\sigma^{\mu\nu} \psi $ & dipole operator & spin torque & CASPEr~\cite{Budker:2013hfa} 
   \\\cline{2-6}
   & axial-vector & $A_\mu' \bar{\psi} \gamma^\mu \gamma^5 \psi$ & minimally coupled  & spin torque & CASPEr~\cite{Budker:2013hfa} 
\\\hline
\,\,\, 2 (?) & tensor & \,$h'_{\mu\nu} T^{\mu\nu} \, (?)$ & gravity-like & grav.\,wave-like & grav.\,wave detectors? \\\hline
  \end{tabular} 
\caption{
The leading couplings of light bosonic dark matter ($\phi$, $a$, $A_\mu '$, and $h'_{\mu\nu}$) to Standard Model fields, and the oscillating physical effects they cause. 
$h$, $G^{\mu\nu}$, $F^{\mu\nu}$, $\psi$, and $T^{\mu\nu}$ represent respectively Standard Model Higgs, gluon, photon, and fermion fields, and the energy momentum tensor, or operators of that form. 
The last column indicates DM searches currently existing or under construction. A star ($\mathbf{\star}$) marks where the searches we propose would lie. 
We note that questions remain about validity, naturalness and allowed interactions in theories of massive spin 2 fields. 
}
\label{table:bosons}
\end{table}

Interestingly there are not too many possibilities for light bosonic field dark matter, and it appears possible to design direct detection experiments to probe them all. 
The state of light bosonic dark matter in the galaxy is well described as a random classical field that:
\begin{itemize}
\item oscillates at an angular frequency equal to the dark matter mass, $m$,
\item is locally coherent over $\sim\!\!10^6$ oscillations (meaning a 1 part in $10^6$ frequency spread),
\item has a local energy density of $\rho \!\sim\! (.04\,\text{eV})^4 \!\sim\! m^2 \phi^2$, where $\phi$ is the boson.  
\end{itemize}
The first is simply true of a non-relativistic field.  The second is due to the fact that the virial velocity of local dark matter is $v\simeq 10^{-3}$ and the field should be coherent over a de Broglie wavelength $1/(m v)$, and thus for a time $\sim\! 1/(m v^2) \simeq 10^6 / m$.  The third assumes any of the standard halo profiles for dark matter, and thus makes a prediction for the amplitude of the field.

As a result of these oscillations, the dark matter acts as a time-dependent source term for the Standard Model operators given in Table~\ref{table:bosons}. This leads to time-dependent signals with the same frequency and phase as the dark matter field.  
The experimental signals can be roughly divided into three categories.  First, the {\it axion-gluon, axion-fermion, vector dipole and axial-vector} couplings produce spin-dependent forces or signals.  Proposals have been made to detect such dark matter, notably though detection of an oscillating neutron electric dipole moment and/or an ``axion wind'' (which exerts a torque on particles' spins)~\cite{Graham:2013gfa, Budker:2013hfa}, which should in principle also be sensitive to vectors coupled through the dipole operator, and axial vectors.  
Independently of their cosmic abundance, these fields can also be searched for in experiments attempting to source and detect a spin-dependent force~\cite{Moody:1984ba, Brown:2010fga, Heckel:2013ina, Terrano:2015}, although the DM searches have sensitivity to significantly smaller couplings. 

The second category of experimental signals comes from the {\it vector-photon mixing and axion-photon coupling}, both of which can produce an effective electromotive force which can drive electric currents. 
The axion-photon coupling results in an EMF in the presence of a background magnetic field, and is being searched for in this way by ADMX~\cite{Asztalos:2009yp}. Vector DM with vector-photon mixing generates an EMF in vacuum inside a shield, which will be searched for, for example, by the recently proposed DM Radio~\cite{Chaudhuri:2014dla}.  
Again, these fields can also be searched for independently of their cosmic abundance, but with less reach in coupling---in this case with experiments that search for transmission of electromagnetic radiation through a shield~\cite{Jaeckel:2007ch, Bahre:2013ywa, Graham:2014sha}.

The third category comes from the {\it Higgs portal,  minimal vector, and gravity-like} couplings, all of which can produce a coherent time-dependent Equivalence Principle (EP) violating force on macroscopic matter, as well as an oscillation of Standard-Model parameters.  {\it Exploiting these effects is the central point of this article}.

\myparagraph{Scalar couplings}
In the case of the scalar, there are a number of potentially important couplings to Standard Model fields.  For example, a linear coupling to quark mass or electron mass operators will produce the time-dependent, EP-violating forces we are interested in.  We focus in this paper on the linear coupling to the Higgs for several reasons:
\begin{itemize}

\item Universality:  The coupling to the Higgs produces (different) couplings to the electron, proton, and neutron masses and thus, in terms of a single parameter, reproduces all of the interesting effects.

\item IR Dominance:  The scalar coupling to the Higgs is the lowest dimension coupling allowed, and therefore any other linear coupling to Standard Model operators will generate this coupling at loop level, whereas the opposite cannot be said (as there are no divergent graphs for higher dimensional operators).  In addition, it is plausible that the UV theory produces this operator with a larger effective coefficient in the IR than all others due to its low dimension.

\item Naturalness:  For a given $\phi$ mass, $m$, and coupling, $b \phi h^\dagger h$, naturalness only requires the coupling to be smaller than the mass, {\it i.e.}  $b\lesssim m$.  Higgs loops also produce a tadpole term for $\phi$ with coefficient of order $b \Lambda^2$, where $\Lambda$ is the scale at which Higgs loops are cut off.  
In the natural regime, $m_\phi \simgt b$, the linear term will produce a vacuum expectation value for $\phi$ of size $\sim\! b \Lambda^2/m^2$, and thus a contribution to the Higgs mass squared smaller than $\Lambda^2$, {\it i.e.} smaller than the direct quantum corrections to the Higgs mass.  Thus, this coupling does not contribute to the naturalness story of the Higgs, and naturalness constraints on $b$ do not depend upon the UV cutoff of the Standard Model.  This is different from all other Standard Model operators $\phi$ could couple to, where the naturalness of its own mass would also depend on the cutoff scale of Standard Model loops. 

\end{itemize}
In addition this direct coupling to the Higgs is interesting because it is the defining coupling of the scalar ``relaxion" in dynamical relaxation solutions of the hierarchy problem \cite{Graham:2015cka}.
While the last constraint makes the Higgs coupling especially interesting, there are individual couplings ({\it e.g.}, to quark and electron masses) which can be both natural and more experimentally accessible (for a sufficiently low cutoff).  We include a direct coupling to the electron mass in our projections as one interesting example.

\myparagraph{Vector couplings}
A minimally coupled vector, which is anomaly free (with respect to mixed anomalies with the Standard Model gauge groups), and allows all Standard Model Yukawa couplings, is any linear combination of hypercharge and baryon minus lepton number ($B-L$).  However, there are more generally a number of allowed vector couplings to the Standard Model:
\begin{itemize}

\item Flavor-blind, anomaly-free:  $(B-L) + Y \rightarrow (B-L) + QED$.  The latter part acts as a dark photon and will be picked up by those searches.  However, bounds on the dark photon are tremendously weaker than bounds on $B-L$, since a dark photon puts a negligible force on neutral matter. 

\item Flavor-blind, anomalous:  $B$ and $L$ are anomalous symmetries, and a field coupled to them would produce a force on matter.  Because of the anomaly, the longitudinal mode of such a vector would become strongly coupled at $\sim\! 4\pi m/g$, where $m$ and $g$ are the mass and coupling (assuming charges of order unity).  This is the highest scale where new fermions could appear to cancel the anomaly (with Yukawa couplings of $4\pi$), and thus we would require $\sim\! 4\pi m/g \simgt 1$\,TeV, which we will see is difficult (but possible) to probe.  Most combinations of $B$ and $L$ will have a similar degree of EP-violation between different isotopes (except for $B$, which has larger suppression and is similar to the scalar).

\item Flavor-dependent, anomaly free: It is easy to construct anomaly free gauge symmetries with flavor-dependent charges. 
However, the required fermion masses in the Standard Model explicitly break such symmetries, which again implies a cutoff at some higher scale. 
For example, if electrons have an axial charge, and muons have the opposite axial charge, the explicit breaking of the gauge symmetry due to the muon Yukawa coupling, $y_\mu$, will induce strong coupling for the longitudinal mode at a scale $\sim\! (4\pi/y_\mu) (m/g)$ (which is a weaker constraint than the anomalous case).  
By far the weakest strong coupling constraint is on a coupling to the difference of lepton flavors, such as $L_e - L_\mu$, which would only be violated by certain neutrino mixing terms.  
Vector bosons couple to fermions of the first generation would produce similar effects to that of $B-L$, whereas couplings to second and third generations only will be much harder to test.

\end{itemize}
Thus, not only is a vector coupled to $B-L$ simplest from a theoretical point of view, but searching for its effects will also cover the effects of many other classes of vectors. Thus, in this paper we will only consider vectors coupled $B-L$ charge.

\section{EP-Violating Signatures}
\label{sec: ep signature}

\subsection{Discussion}

The DM described above causes an equivalence-principle violating acceleration on test bodies, which oscillates in time at the natural frequency of the dark matter (equal to its mass).  
This is of great benefit to experiments searching for this signal, since it is a fundamental frequency of nature unrelated to anything generated by the laboratory or the experiment itself.  
Compare this with searches for static equivalence-principle violating forces in which systematic effects such as gravitational gradients can mimic the signal.  
Handling these systematics is  most difficult for static signals and is much less of a problem in searches for the very distinctive DM signal.  
In section~\ref{sec:torsion-balance}, we consider this explicitly in the concrete setup of torsion pendulums. 
In addition, the narrow frequency spread of the signal could also enable resonant schemes that lead to signal amplification. 

The force on a test body from the dark matter points in a random direction that is uncorrelated with anything in the laboratory and is fixed in the frame of the galaxy over the coherence time.  
This is particularly beneficial in torsion pendulum experiments, which suffer a $\sim\!10^{-3}$ suppression in searches for a vertical force sourced by the Earth, but suffer no such suppression in a DM search. 
In the case of the $B-L$ coupled vector, for example, the force points in the direction of the vector's electric field, which is unknown.  
In the case of the scalar, the force points in the direction of the local gradient in the scalar field (its momentum).  In either case the direction and magnitude of the force change by $\mathcal{O}(1)$ amounts on timescales of $10^6 / m$, the coherence time of the dark matter.  
Of course, since the force is set in the galactic frame, in the earth frame it will also have a daily (and yearly) modulation.  This modulation is useful because it allows us to distinguish the dark matter force from many backgrounds which arise from objects fixed on the earth, for example static gravity gradients.

Given these advantages, what is the optimal way to measure these dark-matter induced accelerations? Current accelerometer technologies are fundamentally sensors of position.  The position of a  test body subject to an acceleration $a$ at a frequency $m$ will oscillate with an amplitude ${\Delta x \sim (a/m^2)}$. These modulations of position can be measured with high precision interferometers (either optical or atomic). Since displacement (as opposed to acceleration) is a relative quantity, this scheme can only measure  relative accelerations, for example  between two test bodies. The dark matter can induce such a relative acceleration in two ways. 
Firstly, if the test bodies are physically separated by some distance, the value of the dark matter field at the two locations will be different, resulting in different forces being exerted on the two objects. 
This can produce a signal in gravitational wave detectors, but with a significant suppression at lower frequencies due to the reliance on the small dark matter gradient ($\sim\!m v$), as was discussed in~\cite{Arvanitaki:2014faa}.

Secondly, if the composition of the two bodies is different, they will experience different accelerations, since the force exerted by the dark matter violates the equivalence principle. 
The size of this effect is suppressed by the degree of EP-violation between the test-body materials. 
However, since the effect is independent of the physical separation of the test bodies, it does not suffer the extra gradient suppression seen in the effect relying on physical separation. 
It also allows experiments to be designed to be insensitive to the time-varying Newtonian gravitational backgrounds, which are known to be significant below $\sim\! 10$ Hz~\cite{Harms:2015zma}.
This is extremely beneficial since the measurable displacement of the text bodies increases as $1/m^2$, making accelerometers generally most sensitive at low frequencies.  

The scalar DM considered in this paper can also produce an oscillations of fundamental ``constants'', such as the electron mass or fine structure constant. 
This approach was considered in~\cite{Arvanitaki:2014faa, VanTilburg:2015oza} (using comparisons of different atomic clocks) and in~\cite{Arvanitaki:2015iga} (using resonant mass detectors), and is considered again here in section~\ref{Sec:Pulsars} (using pulsar timing arrays). 
The proposed atomic clock searches would be sensitive at the lowest possible DM masses, and therefore overlap somewhat with the searches we consider here. 
We find the searches we propose to be generally more powerful, although the projections rely on assumptions about the development of different technologies, and have different relative strengths for different types of coupling.

We now focus on the Higgs-coupled scalar and the $B-L$ coupled vector as two naturally light DM candidates. 
We also consider a scalar coupled to the electron mass operator, as an example of the range of possible non-renormalizable scalar couplings. 
These all induce a time-oscillating, EP-violating force on accelerometers (bulk matter or atoms).  
Searches for this effect constitute DM direct detection experiments, with the signal resulting from the direct contact interaction between the DM and normal matter, similar to axion or gravitational wave detection.

\subsection{\textit{B\,--\,L} vector DM}

We begin with $B-L$ coupled vector dark matter.  Such a light vector field behaves very similarly to a coherent electromagnetic field, except for its mass term and the fact that it also couples to neutrons.  
Its coupling can produce new forces in two distinct and important ways.  
First, it produces a static force between clumps of matter (such as between the Earth and a test mass), and this force is constrained both by tests of the inverse square law and of the EP, depending on the vector mass (see, for example,~\cite{Adelberger:2009zz}).  Second, as we present in this paper, if the vector makes up a portion of the dark matter, it produces an oscillating (time-dependent) EP-violating force directly on matter.  
For a $B-L$ coupling, the dominant EP violating effect is due to the relative neutron fraction of different atoms.  In searches for a static effect, the force arises due vector exchange between source and test masses, and (for distances shorter than the Compton wavelength of the boson) is
\begin{equation}
F_{\text{EP-static}} \simeq g_{B-L}^2 \, \Delta_{B-L}  \left(\frac{A_S-Z_S}{A_S}\right) \frac{M_S M_{A_i}}{m_N^2} \frac{1}{R^2} \, .
\end{equation}
Here $g_{B-L}$ is the coupling strength of the vector, $A_S$ and $Z_S$ are the atomic weight and number of the source, are the masses of the source and test bodies, $R$ is the separation between the source and test bodies,  $M_S$ and $M_A$ are their masses, $m_N$ is the nucleon mass, and $\Delta_{B-L}$ is the degree of EP violation of the two test bodies, given by
\begin{equation}
\Delta_{B-L} = \frac{Z_1}{A_1} - \frac{Z_2}{A_2} \, ,
\label{eq:B-L-EP-violation}
\end{equation}  
where $A_{1,2}$ and $Z_{1,2}$ are the atomic weight and number of the two test bodies. 
Assuming the vectors are the dominant component of dark matter, their density is $\rho_\text{DM} \simeq m_A^2 A^\mu_{B-L} A^{B-L}_\mu \simeq \partial_t A^{B-L}_i \partial_t A^i_{B-L} $ (where we neglect the small spatial gradients).   From the equations of motion, matter couples to $\partial_t A^{B-L}_i \simeq E^{B-L}_i$ as in normal electromagnetism, but with the neutron having the same charge as the proton.  Thus, the magnitude of the EP-violating force is simply
\begin{equation}
F_{\text{EP-DM}} = g_{B-L} E^{B-L} \, \Delta_{B-L} \frac{M_{A_i}}{m_N} \simeq g_{B-L} \sqrt{\rho_\text{DM}} \, \Delta_{B-L}  \frac{M_{A_i}}{m_N}
 \end{equation}
leading to a relative acceleration of
\begin{equation}
\Delta a_{\rm DM} \simeq g \left(\frac{g_{B-L}}{2\times 10^{-11}}\right) \, \Delta_{B-L} \, ,
\label{eq:B-L-acceleration}
\end{equation}
where $g\simeq 9.8\,\text{m}\,\text{s}^{-2}$. 

Comparing the dark-matter induced signal to the static force signal generated by the Earth, we get the simple formula for the ratio of acceleration amplitudes:
\begin{equation}
\frac{a_\text{DM}}{a_\text{static}} \simeq \frac{1}{g_{B-L}} \frac{m_N  \sqrt{\rho_\text{DM}}}{m_{\rm Pl}^2 g} \left(\frac{A_S-Z_S}{A_S}\right) \sim \frac{{1\over 4} \times10^{-27} }{g_{B-L}} \, ,
\end{equation}
The $\mathcal O(10^{-3})$ suppression of the static effect in the E\"ot-Wash experiment means that the dark matter produces a larger acceleration for couplings smaller than $10^{-24} - 10^{-25}$. 
As we see in Fig.~\ref{fig:B-L-plot}, current bounds already push the coupling below this limit, and thus if $B-L$ coupled vectors make up all the dark matter, their effect on the E\"ot-Wash torsion pendulums is larger than the static contribution in the entire remaining low-mass parameter space.

\subsection{Higgs-portal scalar DM}
\label{sec:higgs-portal-scalar-couplings}

We now turn to the more complicated case of a scalar linearly coupled to the Higgs mass operator, $\mathcal L \supset b \phi |H|^2$. 
This translates in a relatively straightforward way to a linear coupling of $\phi$ to all Standard Model fields.  
To compute the effects on matter, we need the couplings to quarks and leptons, and the estimates of the coupling to neutrons, protons and nuclei (as parameterized in \cite{Piazza:2010ye}):
\begin{equation}
\mathcal L \supset \frac{b \phi}{m_h^2}\vev g_{h \psi \psi} \bar{\psi}\psi
\end{equation}
where $\vev=246\, \GeV$ and $m_h = 125\,\GeV$ are the Higgs expectation value and mass.  
The $\psi$ are fermions, and for quarks and leptons, it is simple to compute the leading coupling -- minimizing the Higgs potential and treating $\phi$ as a background field, one gets $g_{hqq}=m_{q}/\vev$ and $g_{h\ell\ell}=m_{\ell}/\vev$ respectively.  
The universal coupling to nucleons has significant uncertainly, and we take $g_{hNN}=200 \, \MeV /\vev \simeq 10^{-3}$ in quoting our bounds.  One can see this as a reasonable estimate from two contributions.  First, if one integrates out the three heavy quarks, top, bottom, and charm, their mass thresholds depend on the scalar, $\phi$, and one can compute contributions to the one-loop strong coupling scale, treating $\phi$ as a background field:
\begin{equation}
\Lambda_{QCD}\rightarrow \Lambda_{QCD} \left( 1 + {2\over 9}{b\phi\over m_h^2}\right) \, .
\end{equation}
 and thus should produce a coupling to nucleons with $g_{h N N}\simeq (2/9) m_N/\vev$, taking the nucleon mass, $m_N$, to scale essentially linearly with $\Lambda_{QCD}$.     The $(2/9)= (b_3^{c} - b_3)/b_3^{c} $, where $b_3^{c}, b_3$ are the beta-function coefficients of QCD just below the charm mass and just above the top mass respectively.  The other significant contribution should come from the strange mass contribution to nucleon masses, according to chiral perturbation theory.  The contribution to the coupling is of the form:
\begin{equation}
 {b\phi\over m_h^2} m_N f_s^N \bar{N} N ,  \:\:\: {\rm with} \:\:\:  f_s^N \equiv \frac{m_s}{m_N}{\partial m_N \over \partial m_s}\, .
\label{eq:f_s}
\end{equation}
Recent lattice calculations, done at the physical point in parameter space, obtains $f_s^N \simeq 0.113\pm 0.053$ \cite{Durr:2015dna}.\footnote{However,  competing calculations extrapolated to the physical point measure disparate values -- see e.g.~\cite{Alarcon:2012nr}, and Figure 2 of~\cite{Durr:2015dna}.}  Finally, the scalar, $\phi$ will have a coupling to photons at one loop by integrating out charged fermions and $W$ bosons.  This contribution is suppressed by a power of the fine structure constant, $\alpha$, and adds only a tiny correction to the scalar coupling to matter.

The force on matter violates EP. 
One reason is that the scalar coupling to neutrons and protons are not precisely proportional to their masses, and differ roughly by the lightest quark masses over the QCD scale, or $\delta \sim \text{few}\,\times 10^{-3}$ \cite{Piazza:2010ye}.  When testing relative forces between different materials, the differential acceleration due to this difference in relative coupling would be proportional to this $\delta$ as well as the factor $\Delta_{B-L} = (Z_1/A_1 - Z_2/A_2)$, where $Z_{1,2}$ and $A_{1,2}$ are the atomic number and mass respectively of the two materials.  In typical experiments testing EP, $\Delta_{B-L}$ is typically a few percent (see for example~\cite{Wagner:2012ui}).  

However, a larger effect should come from the nuclear binding energy differences, especially when comparing light and heavy elements. This is because the binding energy should scale differently with $\Lambda_{QCD}$ and the light quark masses than the nucleon masses do, and thus elements with the largest differences in binding energy per nucleon should produce the largest effect.

Let's see this explicitly.
The mass of an atom is:
\begin{equation}
M_A = N m_n + Z m_p + A \,{\cal E}_B + Z m_e
\end{equation} 
where $N$, $Z$, and $A$ are the neutron number, atomic number, and atomic weight and $m_n$, $m_p$, and $m_e$ are the neutron, proton, and electron masses respectively, and ${\cal E}_B$ is the binding energy per nucleon (and is negative).  
This can be parameterized roughly as:
\begin{eqnarray}
& M_A = A \, m_p + N \delta m_N + A\, {\cal E}_B + Z m_e \\
	\rightarrow & A \, m_n \left(1 + c_N \frac{b \phi}{m_h^2} \right) - Z \delta m_N \left( 1 + c_q \frac{b \phi}{m_h^2} \right) + A  \, {\cal E}_B \left( 1 + c_B \frac{b \phi}{m_h^2} \right) + Z m_e \left(1 +  \frac{b \phi}{m_h^2} \right) \, ,
\end{eqnarray} 
where the second line comes from turning on a background $\phi$. 
Here $\delta m_N = m_n-m_p$,  $c_N \approx 200\MeV/m_N$,  $c_q \approx {\cal O}(1)$, and $c_B$ is expected to be somewhere between $c_N$ and ${\cal O}(1)$.  The $c_q$ coupling is mostly due to the scaling of the nucleon mass difference with the up and down quark mass difference (and a negligible effect from the electromagnetic contribution to the nucleon masses), while $c_B$ is due to the scaling of the binding energy with respect to the QCD scale, strange quark mass, and light quark masses.

Since the coupling of $\phi$ to the atom represents a potential energy for the atom in the presence of the background field, a non-zero gradient for $\phi$ produces a force on atoms.  The acceleration of an element $i$ due to a background $\phi$ is proportional to the linear coupling of $\phi$ to the atom, divided by the atom's mass, or
\begin{equation}
 a_{i} \simeq \nabla\phi \left(c_N - {Z_i\over A_i} \frac{(c_q - c_N) \delta m_N - (1-c_N) m_e}{m_n} + (c_B-c_N)\frac{{\cal E}_{B_i}}{m_n} \right)\frac{b}{m_h^2}
\end{equation}
The second and third terms in parentheses give rise to different accelerations for different materials ({\it i.e.}~EP-violation).  
The effect is typically dominated by the difference in binding energies, so long as $c_B\neq c_N$.  
$c_B$ and $c_N$ have not been precisely measured, but a close cancellation is not expected.   
In particular, $c_N$ is expected to receive a contribution from $m_s$ at the $\mathcal O(0.1)$ level (Eq.~\ref{eq:f_s}), whereas $c_B$ is not. 
The binding energy is also expected to have a dependence on the pion mass, and consequently on the up and down quark masses, that will contribute to $c_B - c_N$ (although some recent lattice calculations have found this contribution to be small~\cite{Beane:2013kca}).  
We therefore take $c_B-c_N = 0.1$ in our calculations, which we assume to be a reasonably conservative estimate.\footnote{Strictly speaking, $c_B$ should depend on the species of atom, but at this point we cannot do better than an $\mathcal O (1)$ estimate.}

In an accelerometer, scalar dark matter will put a direct (time-dependent) EP-violating force on the test masses, which will point in a random direction. 
If the field makes up all the dark matter abundance, then locally 
$\rho_\text{DM} \simeq m_\phi^2 \phi^2 \approx 0.3 \,\GeV/\cm^3 \approx \left( 0.04 \, \eV \right)^4$. 
We can replace the gradient of $\phi$ with the momentum in the field, $\nabla\phi \rightarrow m_\phi v \phi$, where $v\approx 10^{-3}$ is the local dark matter velocity.  
This gives a relative acceleration between two test mass materials of
\begin{equation}
\Delta a_\text{DM} = v \sqrt{\rho_\text{DM}}  \, \frac{b}{m_h^2} \, \Delta_{\phi |H|^2} \, ,
\label{eq:scalar-acceleration}
\end{equation}
where
\begin{equation}
\Delta_{\phi |H|^2}  = \left[ (c_B-c_N)\frac{{\cal E}_{B_1}- {\cal E}_{B_2}}{m_n} - \left({Z_1\over A_1}-{Z_2\over A_2}\right) \frac{(c_q - c_N) \delta m_N - (1-c_N) m_e}{m_n} \right] \, .
\label{eq:scalar-EP-violation}
\end{equation}
We take $(c_q-c_N) \approx (1-c_N) \approx 1$, and $(c_B-c_N) \approx 0.1$. 
For example, with Al/Be test masses, the difference in binding energies dominates (${\cal E}_{B_1} - {\cal E}_{B_2} = 1.87$\,MeV), and $\Delta_{\phi |H|^2}  \approx 2\times 10^{-4}$. On the other hand, for $^{85}$Rb/$^{87}$Rb test masses, the second term is larger, and $\Delta_{\phi |H|^2} \approx 7\times 10^{-6}$. 

The comparison between the DM induced signal and the static effect generated by the Earth is now
\begin{equation}
\frac{\Delta a_\text{DM}}{\Delta a_\text{static}} \simeq v \sqrt{\rho_\text{DM}} \left(c_N \frac{b}{m_h^2} m_{\rm Pl}^2 g\right)^{-1}  \sim \frac{10^{-16} \,\eV}{b}
\end{equation}
Thus, the DM force is stronger when the coupling (and on the naturalness line, the mass) is smaller than $10^{-16}\,\eV$. 
Again, E\"ot-Wash tests of static EP violation receive an extra relative suppression of $\sim\! 10^3$ on top of this estimate, because they are only sensitive to the horizontal force from the center of the Earth~\cite{Wagner:2012ui}. 
Using the numbers given above, the static bounds on a $B-L$ coupled vector of~\cite{Wagner:2012ui} can be converted into bounds on the scalar using $b\equiv 2.6\times 10^{13} g_{B-L} \sqrt{\Delta_{B-L}/\Delta_{\phi|H|^2}}$, where $\Delta_{B-L} = 0.037$ and $\Delta_{\phi|H|^2} = 4\times 10^{-4}$ give the degree of EP violation for the Al/Be test mass combination used to set the limits.

\subsection{Scalar coupled to electron mass}
\label{sec:scalar-coupled-to-electron-mass}

As an example of an alternate coupling of scalar DM, we also consider a dilaton-like coupling only to the electron mass operator,
\begin{equation}
\mathcal L \supset y_{\phi e e}\phi \, \overline e e \, .
\end{equation}
For naturalness the scalar's mass should satisfy (schematically) $m \simgt y_{\phi e e} \Lambda/4\pi$, where $\Lambda$ is the scale at which electron loops are cut off, and we assume $\Lambda \simgt$\,TeV.

The static EP violation sourced by the Earth is related to that in the $B-L$ case by
\begin{equation}
\Delta \vec a \big|_{\phi \overline e e, \, \rm static} = y_{\phi e e}^2 N_{e, \rm source} \left(\frac{Z_1}{A_1 m_N} - \frac{Z_2}{A_2 m_N} \right) \frac{\hat r}{4\pi r^2}
= \frac{y_{\phi e e}^2}{g_{B-L}^2} \frac{N_{e, \rm source}}{N_{n, \rm source}} \times \Delta \vec a \big|_{B-L, \rm \, static} \, .
\end{equation}
Since the Earth has $N_e\approx N_n$, we can therefore just use $y_{\phi e e} \equiv g_{B-L}$ when comparing bounds from static EP tests.

The time-varying EP violation caused by a DM $\phi$ field is related to that in the $B-L$ case by
\begin{equation}
\Delta a \big|_{\phi \overline e e, \, \rm DM} = y_{\phi e e} \nabla \phi \left(\frac{Z_1}{A_1 m_N} - \frac{Z_2}{A_2 m_N} \right)
\approx \frac{y_{\phi e e} v}{g_{B-L}} \times \Delta  a \big|_{B-L, \, \rm DM} \, .
\label{eq:electron-coupled-scalar-acceleration}
\end{equation}
We can therefore use $y_{\phi e e} \equiv g_{B-L}/ v$ when comparing the reach of DM searches, with $v\approx 10^{-3}$.

\section{Experimental Options}
\label{sec: experimental options}

We consider three ways to measure these EP violating forces from dark matter. The methods are distinguished by the nature of the test bodies used to perform the measurement. 
In section~\ref{sec:torsion-balance}, we consider torsion pendulums, with laboratory-scale macroscopic masses whose relative accelerations are measured through optical interferometers. 
In section~\ref{sec:atom-int}, we consider ballistic atoms as test masses, with their relative accelerations measured through atom interferometry. Finally, in section~\ref{sec: LLR} and~\ref{Sec:Pulsars} the test masses are celestial objects---the Moon and pulsars.  In section~\ref{Sec: other bounds} we summarize our projections and discuss other bounds. 
Our results are plotted in Figs.~\ref{fig:B-L-plot},~\ref{fig:scalar-plot} \&~\ref{fig:scalar-plot-2}.

\subsection{Torsion Balances}
\label{sec:torsion-balance}

\subsubsection{Dark matter detection strategy}

\begin{figure}[t]
\begin{center}
\includegraphics[width=0.6\textwidth]{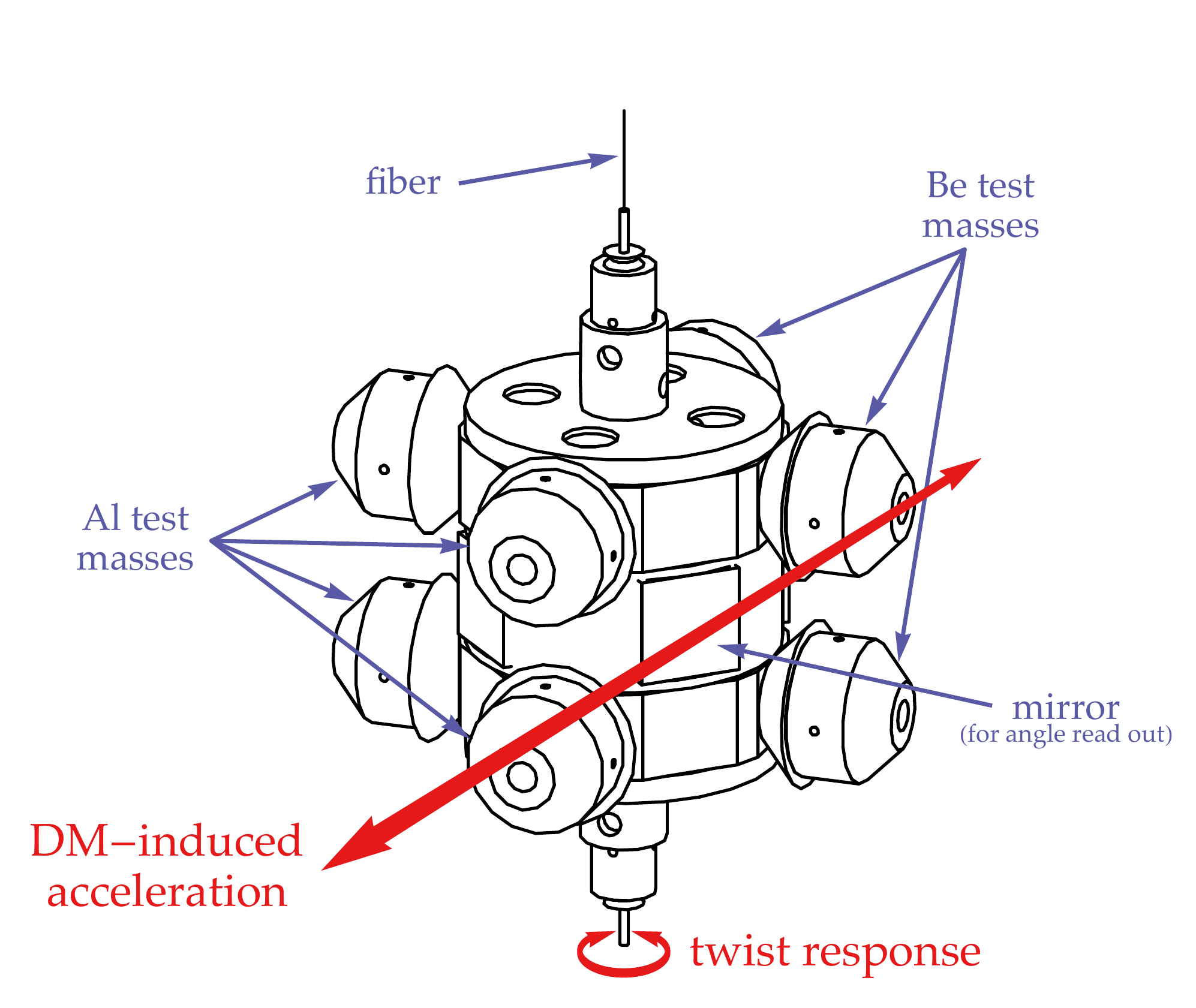}
\caption{A torsion pendulum for dark matter direct detection. 
The dark matter field directly induces an equivalence-principle violating acceleration on the test masses, resulting in an oscillating twist in the the pendulum.  The twist angle is measured by observing the deflection of a laser reflected from the mirror. Adapted from~\cite{Schlamminger:2007ht}.}
\label{fig: EP pendulum}
\end{center}
\end{figure}

Torsion balances are currently the most sensitive instruments for measuring static, equivalence principle (EP) violating forces~\cite{IAU:6911540}.   Here we consider using torsion-balance pendulums to detect the time-oscillating, EP-violating forces induced by light bosonic dark matter, as illustrated in Fig.~\ref{fig: EP pendulum}.

Torsion balances configured for EP tests carry test-bodies of different materials. 
An EP-violating force would apply different accelerations to the different materials, producing a torque, $\tau(t)$, on the pendulum. 
The experiment then consists of carefully monitoring the twist angle $\theta(t)$ of the pendulum to infer $\tau(t)$.

In modern experiments, the entire setup is placed on a turn table rotating at a frequency $f_{\rm tt}$. 
This causes the torque exerted by an otherwise static force to oscillate at this frequency.
The DM signal, by contrast, oscillates at a frequency $f_{\rm DM} = m/2\pi$ (the Compton frequency of the DM), modulated by both the turn-table frequency $f_{\rm tt}$ and the Earth's rotation frequency $f_{\rm \oplus}$. 
Since the DM matter signal does not occur at a frequency associated with either the experiment or the environment, it is naturally distinguishable from most backgrounds.

Torsion balance experiments looking for a static EP-violating force towards the earth (e.g. \cite{Wagner:2012ui, Schlamminger:2007ht}) must deal with any effect that produces a pendulum twist at the turntable rotation frequency, such as turn-table imperfections, temperature gradients, and gravity gradients.  
Consider lab-fixed gravity gradients. These apply a torque on the gravitational moments of the pendulum, and the torque changes direction at frequency $f_{\rm tt}$ as the torsion balance rotates, just like a static EP signal.  
The degree to which one can cancel the gravitational gradients in the lab and the gravitational moments of the pendulum is a limiting factor in these experiments.  
However, the DM signal always has frequency components separated from $f_{\rm tt}$, even for arbitrarily small DM masses, and so only phase-coherent gravity gradient fluctuations that exactly mimic Eq.~(\ref{eq:DM-twist}) affect the measurement.  
This strategy has been effectively used to suppressed the major systematics in searches for sidereally modulated signals in torsion pendulum experiments~\cite{Su:1994gu, Heckel:2008hw}, and it should work just as well in searches for the DM modulated signal.

A second major advantage of the DM search over the static search is the direction of the force.  
The static force sourced by the Earth points in the vertical direction, while a hanging torsion pendulum is sensitive only to horizontal forces.  
Since the Earth's rotation causes the pendulum to hang at a slight angle, the pendulum is sensitive to the static EP force, but only sensitive to at the level of 1 part in $10^3$. 
A DM-induced force, on the other hand, points in a random direction avoids this large suppression.

As a result of the turn-table rotation and the Earth's rotation, the DM signal occurs at 6 distinct frequencies,
\begin{gather}
\tau(t) \sim  \tau_{\rm DM} \sum_i c_i e^{2\pi  f_i t}
\label{eq:DM-twist}
\\
f_i = f_{\rm DM} \pm f_{\rm tt} + \Big\{ \begin{smallmatrix} +f_{\rm \oplus} \\ 0 \\ -f_{\rm \oplus} \end{smallmatrix} \Big\} \, ,
\end{gather}
where $f_{\rm \oplus}$ is the Earth's rotation frequency, as we show in appendix~\ref{sec:signal-in-rotating-balance}.
The overall amplitude $\tau_{\rm DM}$ of the induced torque, and the six $\mathcal O(1)$ coefficients $c_i$, are determined by $\Delta \vec a_{\rm DM}$, the difference in the DM-induced linear acceleration between the two test materials. 
This is given in Eqs.~(\ref{eq:B-L-acceleration}, \ref{eq:scalar-acceleration} \&~\ref{eq:electron-coupled-scalar-acceleration}) for $B$$-$$L$ and scalar DM.
In practice one could then fit the $\theta(t)$ datastream to the functional form from Eq.~\ref{eq:DM-twist}, to fully reconstruct the best fit value of the vector $\Delta \vec a_{\rm DM}$ as a function of $f_{\rm DM}$.  
We elaborate on this briefly in appendix~\ref{sec:signal-in-rotating-balance}, but otherwise we ignore the details of this analysis and simply take
\begin{equation}
|\tau| \approx \tau_{\rm DM} \approx \frac{1}{2} M \, R \, \frac{\Delta a_{\rm DM}}{\sqrt 3} \, ,
\label{eq:torque-approximation}
\end{equation}
where $M$ and $R$ are the combined mass of the test bodies and the arm length of the pendulum. 
The factor $1/\sqrt 3$ relects the fact that only one direction for $\Delta \vec a_{\rm DM}$ contributes to the torque at any instant.

\subsubsection{Noise in torsion balances}

We assume statistical noise sources to be the limiting factor in constraining $\Delta a_{\rm DM}$ with torsion balance data. The statistical uncertainty in torsion balance measurements  come from damping noise in the wire, noise in the angle read out, and external sources of noise such as seismic noise or gravity gradient noise.  
These are described by a noise power $\mathcal P_{\Delta a}(f)$, which controls the uncertainty $\sigma_{\Delta a}$ in the measurement of $\Delta a_{\rm DM}$. 
For an integration time $t_{\rm int}\simlt 10^6/ f_{\rm DM}$, the width of the signal cannot be resolved, and $\sigma_{\Delta a}$ scales as
\begin{equation}
\sigma_{\Delta a}(f) \approx \sqrt{\mathcal P_{\Delta a}(f) \, t_{\rm int}}  
\qquad\qquad (f \simlt 10^6/ t_{\rm int}) \, .
\end{equation}
For longer integration times, when the width of the signal is resolved, $\sigma_{\Delta a}$ can only be improved by performing a bump hunt in $\mathcal P_{\Delta a}(f)$, giving
\begin{equation}
\sigma_{\Delta a}(f) \approx 10^3 \sqrt{\mathcal P_{\Delta a}(f)/ f} \,  (t_{\rm int} f/10^6)^\frac{1}{4}
\qquad\qquad (f \simgt 10^6/ t_{\rm int}) \, .
\end{equation}

$\mathcal P_{\Delta a}(f)$ is related to the noise power in the pendulum twist, $\mathcal P_\theta (f)$, by combining Eq. \ref{eq:torque-approximation} with the pendulum response function given in Eq. \ref{eq:theta-to-torque-power},\begin{gather}
\mathcal P_{\tau} (f) = (2\pi)^4 I^2 \big((f^2-f_0^2 )^2 + f_0^4/Q^2 \big) \mathcal P_\theta (f) \, ,
\label{eq:theta-to-torque-power}
\\
\mathcal P_{\Delta a} (f) \approx \frac{12}{M^2 R^2} \mathcal P_{\tau} (f)
\approx (2\pi)^4 12 R^2  \big((f^2-f_0^2 )^2 + f_0^4/Q^2 \big) \mathcal P_\theta (f) \, ,
\label{eq:torque-to-acc-power}
\end{gather} 
where $2\pi f_0=\sqrt{\kappa/I}$ is the resonant frequency of the pendulum, $I\approx M R^2$ is the moment of inertia, $\kappa$ is the torsional spring constant, and $Q$ is the quality factor. 
The $f^4$ scaling at high frequencies reflects the fact that the twist of the pendulum for a fixed torque drops as $\theta \propto f^{-2}$, since damping becomes irrelevant and the pendulum rotates by the maximum amount possible in a time $1/f$. 

Noise $\mathcal P_{\theta \rm, \, r.o.}$ in the angular read-out system limits the sensitivity to an angular deflection of the pendulum. Converting a flat $\mathcal P_{\theta \rm, \, r.o.}$ spectrum to the corresponding acceleration noise power, using Eqs.~(\ref{eq:theta-to-torque-power}) and~(\ref{eq:torque-to-acc-power}), gives a high-frequency scaling
\begin{equation}
\sqrt{\mathcal P_{\Delta a, \rm \, r.o.} (f)} \approx 3 \times 10^{-9} \, \frac{\text{cm}\, \text{s}^2}{\sqrt\text{Hz}}  \times 
 \bigg(\frac{f}{100 \, \text{mHz}}\bigg)^2 \frac{R}{2\, \text{cm}} \frac{\sqrt{\mathcal P_{\theta \rm, \, r.o.}}}{10^{-9} \, \text{rad}\, \text{Hz}^{-1/2}} \, .
\label{eq:read-out-noise}
\end{equation} 
This is the limiting factor for the DM sensitivity at high frequencies, as seen in Figs.~\ref{fig:B-L-plot} \&~\ref{fig:scalar-plot}.  

Internal damping in the fiber produces thermal noise in the fiber that dominates the noise of the system at low frequencies.  This results in a torque noise $\mathcal P_{\tau \rm, th} (f) = 4 T \kappa/(2\pi f Q)$, and an acceleration noise 
\begin{equation}
\mathcal P_{\Delta a \rm, th} (f) \approx \frac{48 T \kappa}{M^2 R^2 Q} \frac{1}{2\pi f}
\end{equation} 
The spring constant $\kappa$ of the wire scales as $\kappa \propto M^2/L_{\rm wire}$, where $L_{\rm wire}$ is the length of the wire and $M$ is the maximum mass it can support. 
The constant of proportionality depends on the wire materials and how it is fabricated. 
For the 1\,m tungsten wire used in the Eot-Wash setup, this gives a noise scaling 
\begin{equation}
\sqrt{\mathcal P_{\Delta a \rm, th} (f)} \approx 3\times 10^{-10} \, \frac{\text{cm}\, \text{s}^2}{\sqrt\text{Hz}} \times 
 \sqrt{\frac{\rm mHz}{f}} \sqrt{\frac{T}{300\,\text{K}}} \frac{2\, \text{cm}}{R} \sqrt{\frac{5000}{Q}} \, ,
\label{eq: thermal scaling}
\end{equation}  
where we have normalized to the parameters of the current setup~\cite{Schlamminger:2007ht} (see Table~\ref{tab:torsion-params}).  
For a fused silica wire, $\kappa$ and therefore $\mathcal P_{\Delta a \rm, th}$ may be lowered by a factor of $\sim$\,4 compared to tungsten. 
Using fibers with higher $Q$ or cooling the apparatus to lower temperatures would reduce the effect of thermal noise in the measurement, as would increasing the arm length $R$ (while keeping the active-to-passive mass ratio constant).  Torsion pendulums usually have compact (small $R$) designs to minimize couplings to gravity gradients.  If the time dependence of the DM signal does suppress gravitational gradient related systematics sufficiently, a larger pendulum can improve the sensitivity to DM at low frequencies.  
This requires a trade-off with the high-frequency sensitivity, because the effective acceleration noise from the readout system $\mathcal P_{\Delta a \rm, \, r.o.}(f)$ (Eq~\eqref{eq:read-out-noise}) increases with pendulum size.

As shown in Eq.~\ref{eq: thermal scaling}, $ \mathcal P_{\Delta a \rm, th} (f) $ gets worse at low frequencies.  However, rotating the apparatus boosts low signal frequencies to near $f_{\rm tt}$,  so the noise around $f_{\rm tt}$ limits the experiment for all frequencies below $f_{\rm tt}$.  This is why our sensitivity curves are flat at frequencies below $f_{\rm tt}$.  To optimize the experiment the turntable frequency should, if possible, be at the intersection of the angle readout noise limit (dominating at high frequencies) and the thermal noise limit (dominating at low frequencies).  
In practice this can be a challenge to achieve. The values of $f_{\rm tt}$ we use in our sensitivity estimates are intended to be realistic, and are slightly below the optimum values.

\subsubsection{Sensitivity Estimates}

In Figs.~\ref{fig:B-L-plot},~\ref{fig:scalar-plot} \&~\ref{fig:scalar-plot-2} we show estimated sensitivities for 4 possible scenarios:
\begin{enumerate}
\item ``reanalysis'' is the estimated reach from reanalyzing existing data.
\item ``next run'' is based on the continued running of the existing Eot-Wash setup, with the improved autocollimator described in~\cite{Arp:2013yaa} and a higher turn-table frequency. We assume thermal + read-out noise to be the limiting factor. 
\item ``upgrade'' corresponds to a combination of several upgrades which have been demonstrated in the laboratory but have not yet been implemented in EP experiments, so there is some technical risk.
\item ``future'' corresponds to more aggressive upgrades including the use of polypropylene test-bodies, including cooling to liquid Helium temperatures and using silica fibers at the limits of the material properties.  
We also imagine increasing the lever arm of the test-bodies by scaling the pendulum up by a factor of 8.  
This would mean the active and passive mass of the pendulum both increase by 8$^3$ and would require a very sophisticated angle read-out system that reaches the shot noise limit of the system described in  Ref.~\cite{2011OptL...36.1698H}.  
Other strategies to make larger lever arm have various trade-offs, this is just meant to provide an idea of the reach provided by a larger pendulum. 
Overall this represents an aggressive projection with significant technology upgrades. 

\end{enumerate}

The reanalysis estimate is based on the results of Schlamminger et.~al.~\cite{Schlamminger:2007ht}, where a bound was placed on an EP-violating force that is static in the galactic frame (equivalent to the $m\to0$ limit for DM).
The limit was $\tau < 10^{-18}$\,N\,m after $\sim$3 years integration time, using a combined Be/Al test mass of 39\,g and an arm length of 2\,cm, which corresponds to $\Delta a \simlt 3 \times 10^{-13}\,\text{cm}\,\text{s}^{-2}$. 
We take this value as our estimate, although a reanalysis is required, 
since even the smallest DM frequencies would shift the signal away from the frequency looked at in~\cite{Schlamminger:2007ht}.

\vspace{10pt}
\begin{table}[t]
\begin{tabular}{|c|| ccc | c | c c c c c c c|}
\hline
 & \ministack{test-mass}{materials} &  $\Delta_{B-L}$  &  $\Delta_{\phi |H|^2}$  &  \ministack{ $\sqrt P_\theta$ }{[rad/$\sqrt \text{Hz}$]}  &  \ministack{$R$}{[cm]} & \ministack{ $M$ }{[kg]}   &  \ministack{ wire }{material}  &  \ministack{ $\kappa$ }{[erg/rad]}  &  $Q$  &  \ministack{ $T$ }{[K]}   &  \ministack{ $f_{\rm tt}$ }{[mHz]} 
\\ \hline
 ``next run"     &      Al/Be                  & 0.037	& $2\!\times\! 10^{-4}$	& $10^{-9}$	&	2	&	0.04	& tungsten&	0.025	& 5000	& 300	& 10	
\\ ``upgrade" &      Al/Be                    & 0.037	& $2\!\times\! 10^{-4}$	& $10^{-12}$	&	2	&	0.04	& fused silica	&	0.025	& $10^6$	& 300	& 10
\\ ``future" & Be/(C$_3$H$_6$)$_n$ & 0.13	& $1\!\times\! 10^{-4}$	& $10^{-18}$	&	16	&	20.5	& fused silica	&  1640	& $10^8$	& 6		& 100
\\\hline
\end{tabular}
\caption{Experimental setups assumed for the sensitivity curves labelled ``next run'', ``upgrade'' and ``future'' in Figs.~\ref{fig:B-L-plot},~\ref{fig:scalar-plot} \&~\ref{fig:scalar-plot-2}. $\Delta_{B-L}$ and $\Delta_{\phi |H|^2}$ are the degree of EP violation of a $B$$-$$L$ coupled vector and a Higgs-portal coupled scalar, as defined in Eqs.~\ref{eq:B-L-EP-violation} \&~\ref{eq:scalar-EP-violation}, for the given test-mass combination. }
\label{tab:torsion-params}
\end{table}

Table~\ref{tab:torsion-params} shows the assumptions we make for the experimental parameters in the other 3 setups, which we also assume will be limited by thermal and read-out noise as described above. 
Note in particular that it appears possible to improve the angle readout noise of the device significantly.  For example in Ref.~\cite{2011OptL...36.1698H} a sensitivity of $10^{-12} \, \text{rad}\, \text{Hz}^{-\frac{1}{2}}$ was easily achieved with a laser-interferometric setup using an optical lever, so we take this as the ``upgrade'' read-out noise. The shot-noise-limited sensitivity was estimated at $10^{-18} \, \text{rad}\, \text{Hz}^{-\frac{1}{2}}$, so we take this as the ``future'' read-out sensitivity. 
In all cases we assume a 1 year integration time. 

The rotating, cryogenic balance assumed for the ``future'' experiment would be quite a technical feat.  
However, even without rotating the apparatus these sensitivities could still be achieved for frequencies above $\sim$100\,mHz. Below 100\,mHz the DM signal is modulated at the sidereal frequency, so the sensitivity would roll off by a factor of 100 between 100\,mHz and 10\,$\mu$Hz. It is worth noting that the requirements on rotation rate uniformity and tilt-stabilization of the turntable are somewhat less stringent in searches for the DM signal, since the signal is not at the turntable rotation rate.

\subsubsection{Comparison with a linear interferometer}
If the ultimate limit of such a torsion balance experiment would have a laser interferometric angle readout system, one can ask whether this torsion pendulum geometry is an improvement over a linear geometry where the two different masses are hung from their own fibers and the linear rather than differential acceleration between them is measured with a laser interferometer\footnote{This is similar to one arm of LIGO for example except made much shorter and with test masses of differing composition}.  For the linear interferometer the shot noise limit on the displacement sensitivity (in $\frac{\text{m}}{\sqrt{\text{Hz}}}$) is $\sim \frac{\lambda}{\sqrt{N}}$ where $\lambda$ is the laser wavelength and $N$ is the number of photons per second.  The acceleration sensitivity (in $\frac{\text{m}}{\text{s}^2}  \frac{1}{\sqrt{\text{Hz}}} $) to an acceleration $a$ with frequency $f$ is then
\begin{equation}
\label{eqn: linear interferometer sensitivity}
\delta a \sim \frac{\lambda}{\sqrt{N}} f^2.
\end{equation}
By contrast for the torsion pendulum geometry, the sensitivity is parametrically enhanced over this.  The ultimate shot noise limit on angular displacement sensitivity is $ \delta \theta \sim \frac{\lambda}{L \sqrt{N}}$ where $L$ is the length of the optical lever-arm.  The sensitivity to differential linear accelerations of the two proof masses of differing composition is then
\begin{equation}
\label{eqn: torsion interferometer sensitivity}
\delta a \sim \frac{R}{L} \frac{\lambda}{\sqrt{N}} f^2
\end{equation}
where $R$ is the radius of the torsion balance.  Thus it is enhanced over the linear sensitivity Eqn.~\eqref{eqn: linear interferometer sensitivity} by a factor of $\frac{L}{R}$, the optical lever arm over the size of the torsion balance.  So the torsion balance is the optimal geometry for such a signal.

LIGO is an example of a linear interferometer which is very sensitive to accelerations.  The proof masses (mirrors) in LIGO are not made with different materials and so are not designed to search for an EP-violating acceleration.  However there is still a signal in LIGO arising from the time-oscillation of the dark matter field but is suppressed compared to the total acceleration by a factor $\sim mL$ \cite{Arvanitaki:2014faa}.  Initial LIGO only has sensitivity above 30 Hz and Advanced LIGO only has sensitivity above 10 Hz.  LIGO's sensitivity will be roughly comparable to the sensitivity of the ``future" torsion balance curve above these frequencies.  So LIGO will not have sensitivity past current bounds for the scalars, but will have some reach into unexplored parameter space for the $B-L$ gauge boson.

\subsection{Atom Interferometry}
\label{sec:atom-int}

Light pulse atom interferometers~\cite{1991PhRvL..67..181K} have emerged over recent years as precision accelerometers. Currently demonstrated technology is approaching a sensitivity of $\sim\! 10^{-13} \frac{g}{\sqrt{\text{Hz}}}$ at 1 Hz (where $g = 9.8\,\text{m}\text{s}^{-2}$), with sensitivities as high as $\sim\! 10^{-15} \frac{g}{\sqrt{\text{Hz}}}$ achievable in the near term, and perhaps $\sim\! 10^{-17} \frac{g}{\sqrt{\text{Hz}}}$ in the farther future~\cite{Dimopoulos:2006nk, Dimopoulos:2007cj, Dimopoulos:2008sv, Chiow:2011zz, 2013PhRvL.111h3001D}. Experiments to test the equivalence principle using these accelerometers~\cite{Dimopoulos:2006nk, Dimopoulos:2008hx} are presently under construction~\cite{2013PhRvL.111h3001D, 2013PhRvL.111k3002S, 2015PhRvL.114n3004K}. 
These experiments operate by dropping co-located clouds of two different atomic species (e.g. Rb-85 and Rb-87) and measuring the differential acceleration between them. In such experiments, since the atom clouds (the test masses) are in free fall during the course of the measurement, the differential acceleration between them is immune to a variety of noise sources such as seismic vibrations and suspension thermal noise that limit conventional optical interferometers, where these noise sources are introduced due to the need to hold the macroscopic proof masses.  
Atom interferometers face a different set of backgrounds, such as shot noise. 
As with torsion pendulums, searches for static EP violation may ultimately be limited by gravity gradients. 
Again as with torsion pendulums, the time oscillation makes the dark matter signal much less susceptible to these, and there exist protocols to reduce backgrounds down to the shot noise floor~\cite{Graham:2012sy, Dimopoulos:2008sv} for such AC measurements. 

The atom interferometer measures acceleration by sensing the position of the atom clouds at different points in their trajectory. The displacement of the atom clouds scales as $t^2$, the time over which the acceleration is measured. This time is limited by the free fall time $t_f$ of the apparatus and the period $1/m$ of the oscillating dark matter signal. To estimate the sensitivity of the atom interferometers, we will take $t$ to be the smaller of these two times. Since the coherence time of the signal is $\sim$$10^6/m$, it might be possible to improve over this estimate. For masses $m \simgt 1/t_f$, the interferometer could potentially be operated in a resonant mode allowing the displacement to build over several oscillations of the dark matter signal. Similarly,  for $m \simlt 1/t_f$ it might be possible to increase the measurement time by bouncing the atoms many times in the interferometer, effectively increasing the free fall time. Both of these possibilities deserve further experimental consideration, but we do not include them in our sensitivity estimates. 

Under these assumptions, the red curves in Figs.~\ref{fig:B-L-plot},~\ref{fig:scalar-plot} \&~\ref{fig:scalar-plot-2} show our projected sensitivity of atom interferometer searches to an EP-violating dark matter signal. 
From top to bottom, the curves assume interferometers with acceleration sensitivities of  $10^{-13} \frac{g}{\sqrt{\text{Hz}}}$, $10^{-15} \frac{g}{\sqrt{\text{Hz}}}$, and $10^{-17} \frac{g}{\sqrt{\text{Hz}}}$ at 1 Hz.  
In all cases, the sensitivity drops as $1/f^2$ for frequencies $f > 1$\,Hz while remaining flat for frequencies $f < 1$\,Hz, consistent with the above assumption about the measurement time. 
We used Rb-85 and Rb-87 for the atomic species and assumed a total integration time of $10^6$\,s. 
These searches are expected to be more powerful than searches for the static EP-violation caused by the DM field sourced by the Earth (assuming the field to make up all of the DM).

\subsection{Lunar Laser Ranging}
\label{sec: LLR}

Lunar laser ranging (LLR) provides a sensitive search for static EP-violating forces.  Thus it is worth considering whether it would be useful in the search for this type of oscillating, EP-violating force from dark matter.  LLR can detect a new EP violating force from the sun that causes a differential acceleration between the earth and the moon.  Currently the sensitivity is limited by the accuracy of the ranging which is around 1 cm~\cite{Williams:2005rv}.  In order to distinguish the effect of a new force from all the other orbital parameters a fit is performed to the ranging data.  A full statistical model is necessary to precisely calculate the expected reach of LLR for our dark matter signal, which is beyond the scope of the paper.  Here we give a very rough estimate to motivate a more careful search for DM in LLR data.

The force from dark matter will point in a random direction, uncorrelated with the direction to the sun or any other objects in the solar system.  This should aid in discriminating it from the effect of other unknown solar system parameters.  Further, it will oscillate in time with a precise frequency which should also aid in discrimination.  It is likely that if the period of the dark matter is much shorter than a month, the effect on the moon's orbit will average down and so the sensitivity will be reduced.  But for periods longer than a month (dark matter masses below $\sim 10^{-21}$ eV) the effect should be unsuppressed.  If we use the relative (EP-violating) acceleration of the moon and earth caused by $B-L$ vector dark matter from Eq.~\eqref{eq:B-L-acceleration} and then multiply by $(1 \, \text{month})^2$ we find a rough estimate for the extra distance apart the two are separated by after a month.  Requiring this be larger than 1 cm yields a very rough estimate for the sensitivity of LLR to this type of dark matter.  This gives a reach in the $B-L$ coupling which is down to $g \sim 5 \times 10^{-26}$ in the mass range below about $\sim 10^{-21}$ eV, about an order of magnitude below current bounds.  Of course the real estimate may be better than this since there are already many months worth of data, and the sharp peak in the frequency spectrum of the signal may allow noise to be removed.  It would thus be very interesting to reanalyze the LLR data to search for this dark matter signal.

\subsection{Pulsar Timing Arrays}
\label{Sec:Pulsars}

Observational constraints permit the mass of the dark matter to be as low as $m \!\sim\! 10^{-22}$\,eV, leading to oscillating effects with periods as long as $\sim$\,year. 
Celestial test bodies offer a powerful way to probe signals at these ultra-low frequencies. In particular, leveraging the exceptional stability of pulsars, timing arrays that measure differences in the arrival times of signals from a number of pulsars have been proposed as a way to search for gravitational waves in the $10^{-8}-10^{-6}$\,Hz frequency band. 
These pulsar timing arrays are automatically sensitive to the ultra-light bosonic dark matter considered in this paper. 

One possible signal, considered in~\cite{Khmelnitsky:2013lxt, Porayko:2014rfa}, arises due to the purely gravitational effect the oscillating dark matter field has on the metric. 
However, much larger effects can arise due to the direct couplings of the dark matter field to ordinary matter. 
In the case of the scalar dark matter, the strongest effect arises directly in the timing measurements, which are made using atomic clocks. 
The oscillation of the scalar field causes the electron mass to oscillate, as discussed in Section \ref{sec: theory}.  
This causes the timing of all terrestrial clocks to oscillate with the same fractional magnitude, since atomic clock frequencies are set by the spacings between atomic energy levels which are proportional to the electron mass.  
Of course, the pulsars ``clocks" are also affected by the scalar field oscillations.  
For example the neutron mass also changes and this will affect the moment of inertia of the pulsar and hence its spin rate.  
However, the pulsars being observed are far enough from the earth that they are all well beyond the Compton wavelength of even the lightest possible dark matter candidate.  
Therefore the observed effect on the timing for each pulsar is $\mathcal O(1)$ different from the effect on terrestrial atomic clocks, and also $\mathcal O(1)$ different from the effect on every other pulsar (and does not rely on the EP violation of the force).  
Averaging many pulsars together then essentially provides a stable definition of time where the dark matter effect averages to zero (or more precisely is reduced by the square root of the number of pulsars).  
This definition of time is then compared to terrestrial atomic clocks which have oscillations in their timing caused by the dark matter.  
This is conceptually similar to a comparison between different atomic clock~\cite{Arvanitaki:2014faa, VanTilburg:2015oza}, although here the pulsars provide a standard of time to compare to which essentially does not oscillate with the dark matter.  
This gives a timing residual of
\begin{gather}
\Delta t = t_{\rm clock} - t = \int d t \frac{\omega_{\rm clock}(t)}{\omega_{\rm clock, 0}} - t \approx  \frac{1}{m} \left|\frac{\delta m_e}{m_e} \right| \sin m t \, .
\end{gather}
For the Higgs-portal scalar coupling, $|\delta m_e/m_e | \approx b \sqrt {\rho_{\rm DM}}/(m \, m_h^2)$, while for a scalar coupled to the electron mass $|\delta m_e/m_e | \approx g_{\phi e e} \sqrt {\rho_{\rm DM}}/(m \,m_e)$.
This signal is observable in pulsar timing arrays in a similar way to a gravitational wave.
 
We estimate the sensitivity to scalar dark matter of the existing European Pulsar Timing Array (EPTA), and of 10 years of running of the upcoming Square Kilometer Array (SKA)~\cite{Carilli:2004nx}, by comparing the maximum timing residual to the timing sensitivity of the array. 
From~\cite{Moore:2014lga}, this sensitivity is $\Delta t \approx \sigma_t/\sqrt{N_p N_m}$, where $\sigma_t \!\sim\! 100 \,\text{ns}$ ($30 \,\text{ns}$), $N_p = 5$ (50) is the number of pulsars, and $N_m = 10 \,\text{yr} / 2 \,\text{weeks}$ is the number of measurements per pulsar.  
(This is approximately equivalent to equating the maximum fractional change of $m_e$ to the gravitational wave strain sensitivity estimated in~\cite{Moore:2014eua}.) 
We show our estimates of these sensitivities in Figs.~\ref{fig:scalar-plot} \&~\ref{fig:scalar-plot-2}.
 
\begin{figure}[t]
\begin{center}
\includegraphics[width=0.8\textwidth]{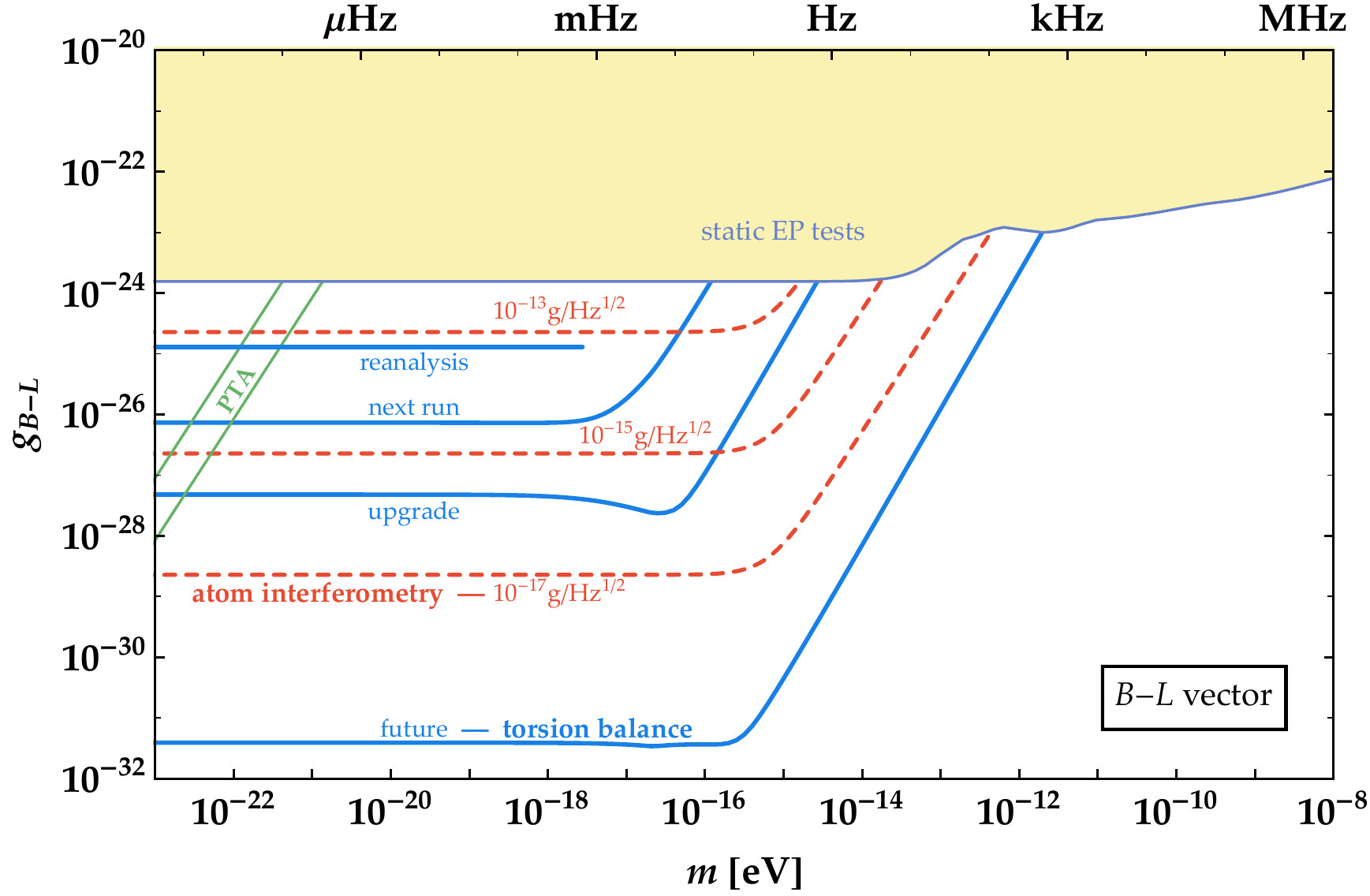}
\caption{
Estimated reach of searches for $B$$-$$L$-coupled vector DM.
Solid blue curves correspond to the torsion pendulum setups discussed in Sec.~\ref{sec:torsion-balance}. 
Dashed red curves correspond to atom interferometers with acceleration sensitivities of $10^{-13}$, $10^{-15}$, and 
$10^{-18}\,\text{g}\,\text{Hz}^{-1/2}$, as discussed in Sec.~\ref{sec:atom-int}.
Thin green curves correspond to the existing EPTA (upper) and upcoming SKA (lower) pulsar timing arrays, as discussed in Sec.~\ref{Sec:Pulsars}. 
The shaded region show bounds from static EP tests using torsion pendulums~\cite{Wagner:2012ui}. 
We have not shown estimates for the sensitivity of lunar laser ranging to DM, but even existing data is potentially powerful (see Sec.~\ref{sec: LLR}).
}
\label{fig:B-L-plot}
\end{center}
\end{figure}

\begin{figure}[t]
\begin{center}
\includegraphics[width=0.8\textwidth]{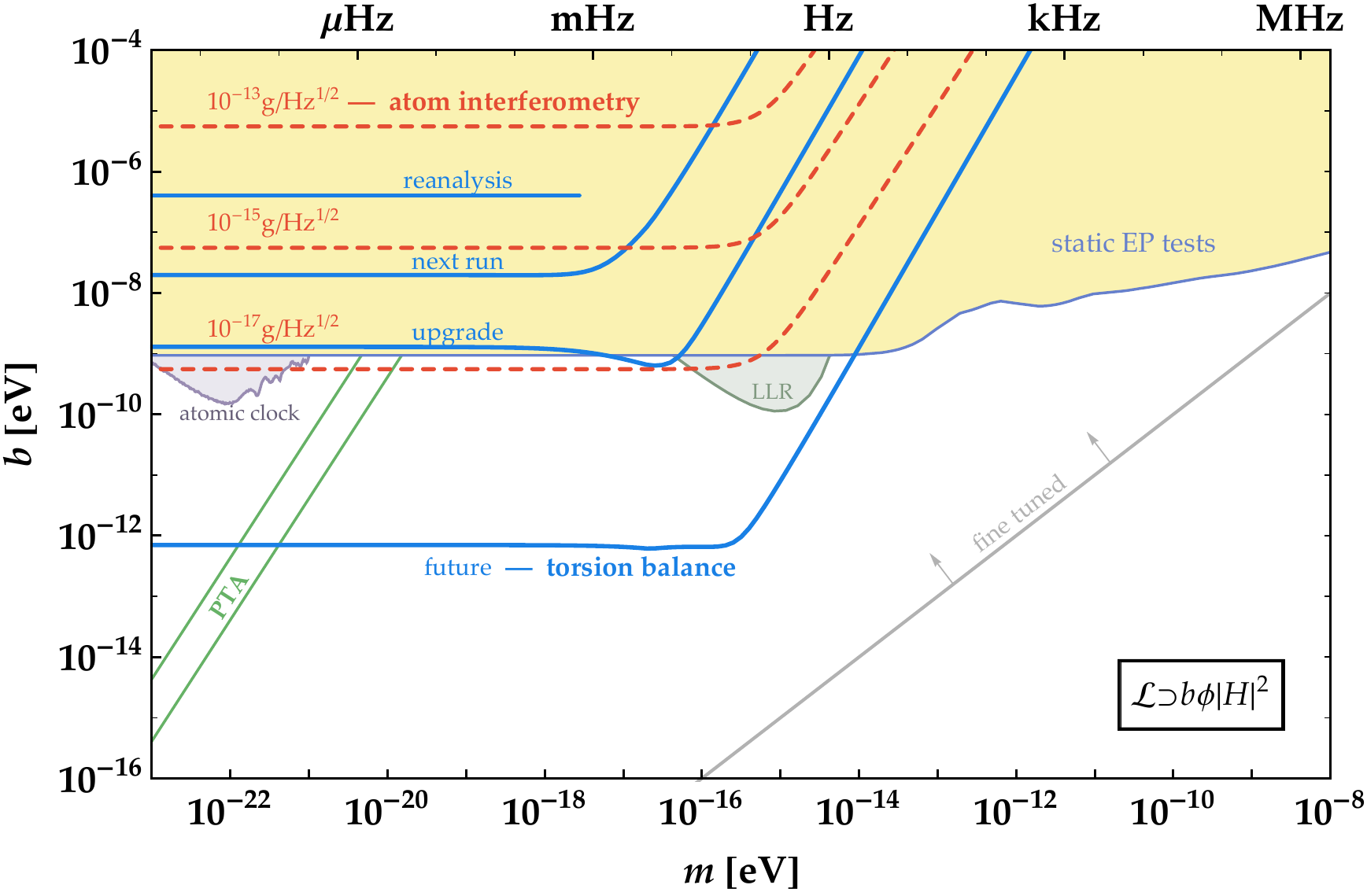}
\caption{
Estimated reach of searches for scalar DM coupled through the Higgs portal, $\mathcal L \supset b \phi |H|^2$. 
The blue, red and green curves and the yellow shaded region are as in Fig.~\ref{fig:B-L-plot}.
Also shown are regions excluded by tests of GR using Lunar Laser Ranging~\cite{Schlamminger:2007ht, Turyshev:2006gm},
and a search for scalar DM in atomic clock data~\cite{VanTilburg:2015oza}. 
Above the gray line the scalar mass is fine tuned. 
As discussed in Sec.~\ref{sec:higgs-portal-scalar-couplings}, there is an $\mathcal O(1)$ theory uncertainty in the degree of EP violation of the coupling to nuclei, which affects both the static EP-test bounds and the projections for atom interferometer and torsion balance tests. 
}
\label{fig:scalar-plot}
\end{center}
\end{figure}

\begin{figure}[t]
\begin{center}
\includegraphics[width=0.8\textwidth]{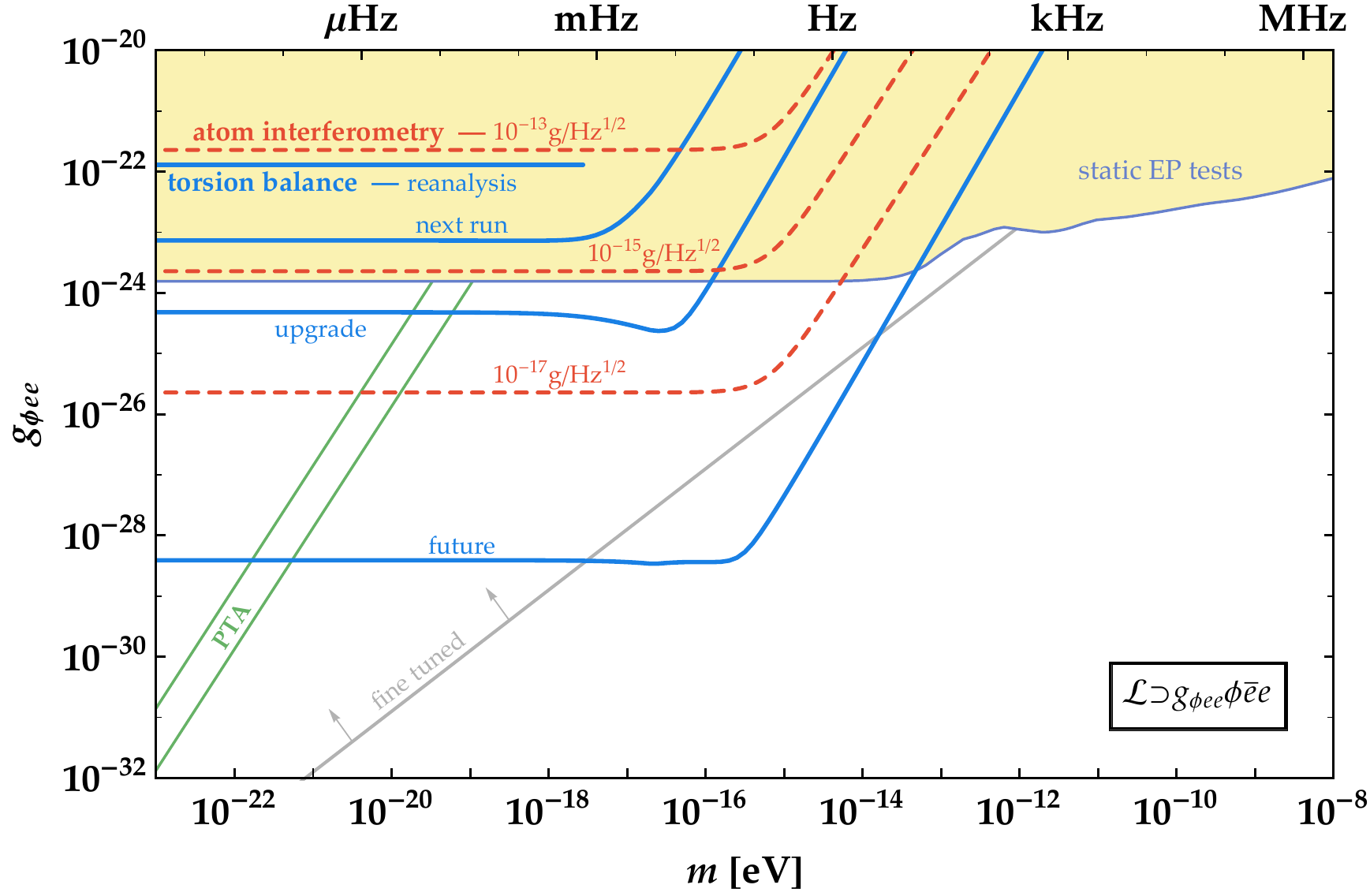}
\caption{
Estimated reach of searches for scalar DM coupled through $\mathcal L \supset g_{\phi e e} \phi \overline e e$. 
The blue, red and green curves and the yellow shaded region are as in Fig.~\ref{fig:B-L-plot}.
Above the gray line the scalar mass is fine tuned if the cutoff $\Lambda$ is above $\sim\!\text{TeV}$.
}
\label{fig:scalar-plot-2}
\end{center}
\end{figure}

Dark matter also gives rise to a timing residual due to the force it exerts on matter. 
This results in a physical displacement of both the Earth and pulsars, which as before are not in phase with each other due to the large separation lengths, resulting in an observed timing residual (again this does not rely on the EP-violation of the force, which anyway is relatively large).
For the Higgs-coupled scalar, this effect is $\mathcal O(10^{-4})$ smaller than the dilation effect due to the velocity suppression of the force, and the smaller fractional coupling of the scalar to the nucleon mass than the electron mass. 
For the $B$$-$$L$ coupled vector it gives the dominant effect, resulting in a timing residual of
\begin{equation}
\Delta t = \Delta x = \frac{a}{m^2} \approx \frac{g_{B-L} \sqrt{\rho_{\rm DM}}}{2 \sqrt 3 m_n \, m^2} \,
\end{equation}
where the factor of 2 accounts for the fact that the Earth is $\sim$50\% neutrons, and the factor of $\sqrt 3$ accounts for the fact that we only observe displacements along the line of sight to a given pulsar. Comparing to the timing sensitivity discussed above gives the estimated sensitivity shown in Fig.~\ref{fig:B-L-plot}.

Both the time-dilation signal and the acceleration signal are similar to a gravitational wave signal in a pulsar timing array, with the key difference that these signals are not tensor in nature (they are scalar and vector in nature respectively).  A gravitational wave causes contraction along one axis and expansion in a perpendicular axis, and thus the timing changes from pulsars depend on the direction to the pulsar in this tensorial manner.  However in the case of scalar dark matter, the timing change is describable just as a change to terrestrial atomic clocks and is therefore independent of the direction to the comparison pulsar.  For an acceleration signal, the earth accelerates in a particular direction.  Thus the timing changes to pulsars in that direction have one sign, the timing changes to pulsars in the opposite direction along the same axis have a different sign, and there is no timing change to pulsars in the two perpendicular directions.  While such scalar and vector signals are certainly visible to pulsar timing arrays, searching for them likely requires a slightly different analysis than the gravitational wave analysis.  It would thus be interesting for the pulsar timing arrays to undertake searches for scalar and vector signals.

\subsection{Projected sensitivities}
\label{Sec: other bounds}

In Figs.~\ref{fig:B-L-plot},~\ref{fig:scalar-plot} \&~\ref{fig:scalar-plot-2} we show our estimated sensitivities for the various torsion pendulum, atom interferometer, and pulsar timing array setups discussed above. 
We emphasize that the upper curves for the torsion pendulum and pulsar timing array correspond to possible reanalyses of existing data. 
In all three cases the lowest curves are expected to be reachable in the timescale of a decade.

We also show several existing constraints. 
The yellow shaded regions show bounds from static EP tests using torsion pendulums~\cite{Wagner:2012ui}. 
We converted these bounds from the $B$$-$$L$ case into the scalar cases according the formulae given in sections~\ref{sec:higgs-portal-scalar-couplings} \&~\ref{sec:scalar-coupled-to-electron-mass}. 
The green shaded region in Fig.~\ref{fig:scalar-plot} labelled LLR shows the bound from tests of the gravitational inverse-square law using lunar laser ranging~\cite{Schlamminger:2007ht, Turyshev:2006gm}. In Figs.~\ref{fig:B-L-plot} \&~\ref{fig:scalar-plot-2} these bounds are weaker than the static EP bounds and so do not appear. 
The static EP and lunar laser ranging bounds arise due to the light field being sourced by the Earth, and are independent of the field's contribution to dark matter.  

The light purple shaded region in Fig.~\ref{fig:scalar-plot} shows the bound from a search for scalar dark matter in atomic clock data~\cite{VanTilburg:2015oza} appearing as oscillating change to $\alpha$. We converted these into bounds on $b$ assuming a coupling $\mathcal L \supset (g_{h \gamma \gamma} b/ m_h^2) \phi F_{\mu \nu} F^{\mu \nu} $, with $g_{h \gamma \gamma} \approx \alpha / 8\pi$~\cite{Piazza:2010ye}. Dedicated atomic clock searches may have significantly increased sensitivity to scalar dark matter in the future~\cite{Arvanitaki:2014faa}. 
The gray lines in Figs.~\ref{fig:scalar-plot} \&~\ref{fig:scalar-plot-2} mark the edge of the natural region (assuming a cutoff above $\sim\!\text{TeV}$ in the electron-coupled case); the parameter space to the left requires fine tuning.

As can be seen from the plots, there is significant potential to probe untested DM parameter space, most dramatically in the case of $B-L$ vector DM or a scalar DM coupled to the electron mass.

\section{Conclusions}

Ultra-weakly coupled, light bosonic particles emerge in a number of theories of physics beyond the Standard Model, and are an important target for experimental searches.  They may play an important role in resolving major problems in particle physics such as the strong CP problem, the electroweak hierarchy problem~\cite{Graham:2015cka}, and possibly the cosmological constant problem~\cite{Abbott:1984qf}. In these approaches, the solutions to these outstanding problems are not the result of high energy physics but rather the product of low energy dynamics.  If these fields exist, it is natural for them to be dark matter. 
It is important to pursue the experimental signatures of these alternative possibilities.
The ultralight mass range, where dark matter behaves as a coherent oscillating classical field, spans over twenty orders of magnitude in mass from $\sim\!10^{-22}$\,eV to $\sim$\,eV.  
Here, we have proposed probes of the lighter half of this parameter space, using sensitive measurements of time-dependent, equivalence principle (EP)-violating accelerations.  Existing accelerometers are already sensitive to new parameter space, and we project that near future technology will improve bounds up to ten orders of magnitude in coupling strength.

Quantum field theory restricts the ways in which  a particle can be naturally light to a small number of leading possibilities, such as derivative interactions (such as axions), kinetic mixing with electromagnetism, minimal gauge couplings, and the higgs portal. 
There are a variety of experiments that source and detect these particles in the laboratory, such as experiments that search for spin dependent forces (for axions~\cite{Moody:1984ba, Brown:2010fga, Heckel:2013ina, Terrano:2015}), the transmission of electromagnetic radiation through a shield (for kinetically mixed hidden photons~\cite{Jaeckel:2007ch, Graham:2014sha}) and experiments that search for new short distance/equivalence-principle violating forces between test bodies~\cite{Adelberger:2013jwa, Chiaverini:2002cb, Dimopoulos:2006nk}. 
The other option besides experiments that both source and detect these particles is to search for an existing cosmic abundance of such particles. There have been a handful of ideas of how to discover these new fields if they make up the dark matter.  

The effects discussed in~\myref{Section}{sec: theory} encompass {\it all} the experimental avenues to directly search for the interactions of ultra-light dark matter. 
A variety of experimental avenues  to search for the time dependent signals caused by a cosmic abundance of axions and hidden photons are currently being pursued, while there are no experiments designed to search for the time varying accelerations caused by $B-L$ gauge bosons and scalars (such a relaxion). 
We can search for ultralight dark matter through its effects on photons, electrons and nucleons. 
Presently, microwave cavities~\cite{Asztalos:2009yp} (and potentially, LC resonators~\cite{Scott:talk, Sikivie:2013laa}) are used to search for axions through their mixing with photons.  
Scalar dark matter coupling to photons or fermions can cause oscillations of fundamental parameters of the Standard Model, which may be searched for using atomic clocks~\cite{Arvanitaki:2014faa, VanTilburg:2015oza} or possibly resonant bars~\cite{Arvanitaki:2015iga}. 
The effects of dark matter on fermions such as electrons and nucleons fall into two broad categories: the dark matter can lead to precession of their spins  (the target of the CASPEr experiment~\cite{Graham:2013gfa, Budker:2013hfa}) or lead to forces acting on them. Of the many possible kinds of forces, there is one particular combination that is aligned with electromagnetism (the hidden photon) and this force can be searched for using LC resonator experiments~\cite{Chaudhuri:2014dla}. 
Any other type of force will  result in an equivalence-principle violating acceleration of electrically neutral atoms. 
The phenomenology will thus be similar to the concepts discussed in this paper.

Our experimental proposal relies on the fact that a cosmic abundance ultralight vectors or scalars may generically exert a direct, time-oscillating, EP-violating force on normal matter. 
This gives rise to accelerations that can be measured by observing the resulting displacement of test bodies.  
Since this displacement scales as $m^{-2}$, an acceleration signal that has a fixed magnitude (such as the signal from dark matter) leads to a larger measurable displacement signal at lower frequencies/masses. 
Since accelerometers appear to be ill suited to search for high frequency signals, it would be interesting to develop alternate protocols that can probe these high frequency regions. 
The EP-violating nature of the dark matter signal enables its detection in compact laboratory setups, since there is a no need for a long baseline. 
This suppresses backgrounds such as time varying gravity gradient noise and seismic vibration noise that normally plague long baseline accelerometer setups such as terrestrial gravitational wave detectors. 
Removing these backgrounds makes it possible to search for low frequency signals, where accelerometer sensitivity is maximized. 
Thus, accelerometers such as torsion pendulums and atom interferometers are optimal ways to directly detect such light dark matter over 10 orders of magnitude in mass from $\sim\! 10^{-8}$\,Hz to kHz.
Our present proposals help establish a full experimental program to discover light dark matter.

\section*{Acknowledgements}
We would like to thank E.~Adelberger, S.~Dimopoulos, R.~Harnik, J.~Hogan, A.~Sushkov, K.~Van Tilburg.
PWG acknowledges the support of NSF grant PHY-1316706, DOE Early Career Award DE-SC0012012, and the W.M.~Keck Foundation.  
SR was supported in part by the NSF under grants PHY-1417295 and PHY-1507160, the Simons Foundation Award 378243. 
DEK acknowledges the support of NSF grant PHY-1214000.
This work was supported in part by the Heising-Simons Foundation grants 2015-037 and 2015-038.

\appendix

\section{DM signal in a rotating torsion-balance}
\label{sec:signal-in-rotating-balance}

In this appendix we derive and discuss the form of the signal in a rotating pendulum arising due to a DM-induce acceleration that is fixed in the galactic rest frame. The result is of importance for discriminating a DM signal from backgrounds: the signal has a distinctive pattern, with 6 different frequencies with correlated phases and amplitudes.

\begin{itemize}

\item Let $\{ \hat x_\oplus, \hat y_\oplus, \hat z_\oplus \}$ define coordinates for an observer standing on the Earth, chosen so that $\hat z_\oplus$ is the axis of the Earth's rotation, and $\hat y_\oplus$ is east. Since these coordinates rotate with the Earth, a vector $\vec E$ that is not rotating with the Earth has the form
\begin{align}
\vec E(t) &= \, E_1(t) (\cos(2\pi f_\oplus t) \hat x_\oplus - \sin(2\pi f_\oplus t) \hat y_\oplus)
\nonumber \\
& \,\,\, + E_2(t) (\sin(2\pi f_\oplus t) \hat x_\oplus + \cos(2\pi f_\oplus t) \hat y_\oplus)
\nonumber \\
& \,\,\, + E_3(t)  \hat z_\oplus \, ,
\end{align}
where $f_\oplus = 4.84 \times 10^{-6} \, \text{Hz}$ is the Earth's rotation frequency.

\item In a lab at latitude $\alpha$, let $\{ \hat x_{\rm lab}, \hat y_{\rm lab}, \hat z_{\rm lab} \}$ point south, east, and upwards respectively. 
Then then $\hat y_{\rm lab} = \hat y_\oplus$ and $\hat x_{\rm lab} = \sin \alpha \, \hat x_\oplus - \cos \alpha \, \hat z_\oplus$. 

\item A torsion pendulum on a horizontal rotating turn table feels a torque from the azimuthal component of $\vec E$, i.e. the component along the direction $\hat \theta = \sin(2\pi f_{\rm tt} t) \hat x_{\rm lab} - \cos(2\pi f_{\rm tt} t) \hat y_{\rm lab}$, where $f_{\rm tt}$ is the turn-table frequency.
This is given by
\begin{align}
E_\theta(t) &= E_1(t) ( \sin \alpha \cos(2\pi f_\oplus t) \sin(2\pi f_{\rm tt} t) + \sin(2\pi f_\oplus t) \cos(2\pi f_{\rm tt} t))
\nonumber \\
& \,\,\, + E_2(t) ( \sin \alpha \sin(2\pi f_\oplus t) \sin(2\pi f_{\rm tt} t) - \cos(2\pi f_\oplus t) \cos(2\pi f_{\rm tt} t))
\nonumber \\
& \,\,\, - E_3(t) \cos \alpha \sin(2\pi f_{\rm tt} t) \, .
\end{align}

\item If $\vec E(t)$ has frequency $f_{\rm DM}$ in galactic frame, there will therefore be signals in the torsion pendulum at 6 different frequencies:
\begin{equation}
f_{\rm obs} = \Big| f_{\rm DM} \pm f_{\rm tt}  + \Big\{ \begin{smallmatrix} +f_\oplus \\ 0 \\ -f_\oplus \end{smallmatrix} \Big\}  \Big|
\end{equation}

\end{itemize}

As discussed in Sec.~\ref{sec:torsion-balance}, this splitting pattern is a distinctive signature of the extra-terrestrial origin of the signal, and so will be a useful test of a DM signal. It can be resolved as long as $f_\oplus \simgt \Delta f_{\rm DM} \approx 10^{-6} f_{\rm DM}$, i.e. $f_{\rm DM} \simlt 0.5 \, \text{Hz}$, and as long as the integration time is longer than $\sim 1$\,day (although it will compete with systematics which modulate daily). 
It will also reduce the backgrounds for the for extremely low mass DM, since a static gravity gradient or a turn-table imperfection appears as signals at frequency $f_{\rm tt}$, while a DM signal has components shifted from this by at least $f_\oplus$.

Since there are 6 signal frequencies (12 real parameters) but only 3 components of $\vec E$ (6 real parameters), there are correlations between the amplitudes and phases of the different frequency components. In principle this allows the full (complex) vector $\vec E$ to be determined, and the remaining constraints provide further checks of the signal. For $B-L$ vector DM, $E_1$, $E_2$ and $E_3$ are expected to be uncorrelated and have the same rms amplitude. For scalar DM, $\vec E$ is replaced with $\vec \nabla \phi$, which is expected to point on average in the direction of the Earth's motion. This difference should ultimately allow the two cases to be distinguished.

\bibliography{PRD_submit_v1}
\bibliographystyle{utphys}

\end{document}